# PGDM: Physically guided diffusion model for land surface temperature downscaling

Huanyu Zhang[a,c,d], Bo-Hui Tang[b,e,f,a,*], Tian Hu[d,**], Yun Jiang[a], Zhao-Liang Li[a,g]

[a] State Key Laboratory of Resources and Environmental Information System, Institute of Geographic Sciences and Natural Resources Research, Chinese Academy of Sciences, Beijing 100101, China

[b] Faculty of Land Resource Engineering, Kunming University of Science and Technology, Kunming 650093, China

[c] College of Resources and Environment, University of Chinese Academy of Sciences, Beijing 100049, China

[d] Luxembourg Institute of Science and Technology, Belvaux 4362, Luxembourg

[e] Yunnan Key Laboratory of Quantitative Remote Sensing, Kunming 650093, China

[f] Yunnan International Joint Laboratory for Integrated Sky-Ground Intelligent Monitoring of Mountain Hazards, Kunming 650093, China

[g] State Key Laboratory of Efficient Utilization of Arid and Semi-arid Arable Land in Northern China, Institute of Agricultural Resources and Regional Planning, Chinese Academy of Agricultural Sciences, Beijing 100081, China

[*] Corresponding author at: Faculty of Land Resource Engineering, Kunming University of Science and Technology, Kunming, 650093, China.

[**] Corresponding author at: Luxembourg Institute of Science and Technology, Belvaux 4362, Luxembourg.

*E-mail addresses:* tangbh@kust.edu.cn (B.-H. Tang), tian.hu@list.lu (T. Hu)

**Abstract:** Land surface temperature (LST) is a fundamental parameter in thermal infrared remote sensing, while current LST products are often constrained by the trade-off between spatial and temporal resolutions. To mitigate this limitation, numerous studies have been conducted to enhance the resolutions of LST data, with a particular emphasis on the spatial dimension (commonly known as LST downscaling). Nevertheless, a comprehensive benchmark dataset tailored for this task remains scarce. In addition, existing downscaling models face challenges related to accuracy, practical usability, and the capability to self-evaluate their uncertainties. To overcome these challenges, this study first compiled three representative datasets, including one dataset over mainland China containing 22,909 image patches for model training and evaluation, as well as two datasets covering 40 heterogeneous regions worldwide for external evaluation. Subsequently, grounded in the surface energy balance (SEB)-based geophysical reasoning, we proposed the physically guided diffusion model (PGDM) for LST downscaling. In this framework, the downscaling task was formulated as an inference problem, aiming to sample from the posterior distribution of high-spatial-resolution (HR) LST conditioned on low-spatial-resolution (LR) LST observations and a suite of HR geophysical priors. Comprehensive evaluations demonstrate the effectiveness of PGDM, which generates high-quality downscaling results and outperforms existing representative interpolation, kernel-driven, hybrid, and deep learning approaches. Finally, by exploiting the inherent stochasticity of PGDM, the scene-level standard deviation of multiple generations was computed, revealing a strong positive linear correlation with the actual downscaling error. This property enables PGDM to self-assess its downscaling uncertainty, constituting an additional key advantage over conventional deterministic downscaling models. The codes and data will be released at https://github.com/cas222huan/PGDM.

**Keywords:** Land surface temperature downscaling, Geophysical reasoning, Diffusion model, Uncertainty indicator, Benchmark dataset

# 1. Introduction

Land surface temperature (LST) is a key biophysical variable in terrestrial ecosystems, and plays an essential role in land surface hydrological and thermal processes (Z. Li et al., 2023; Li et al., 2013). Given its importance in surface energy balance (SEB) and land-atmosphere interactions, LST has been widely used in various downstream applications, such as estimating surface longwave radiation and evapotranspiration (Bhattarai et al., 2019; T. Hu et al., 2023; Liang et al., 2019; Qin et al., 2020; Raoufi and Beighley, 2017), monitoring agricultural drought (Anderson et al., 2016; Hu et al., 2020; Jia et al., 2025; Son et al., 2012), and analyzing urban thermal environments (Fu and Weng, 2016; He et al., 2024; Li et al., 2012; Wang et al., 2019; Zhan et al., 2024).

Although remote sensing remains the most effective approach for LST estimation, the trade-off between spatial and temporal resolutions imposed by sensor fabrication limitations severely constrains the practical applications of LST products (Dong et al., 2020; Mao et al., 2021; Pu and Bonafoni, 2023). To achieve LST retrieval at a high spatiotemporal resolution, numerous studies have been conducted to postprocess remote sensing LST products, with a particular emphasis on the spatial dimension. The methods employed to enhance the spatial resolution of LST, referred to as LST downscaling or sharpening techniques (Yoo et al., 2020; Zhan et al., 2013), can be broadly categorized into six major groups, including kernel-driven models, weight-function fusion methods, thermal unmixing methods, physical methods, deep learning models, and hybrid methods.

The kernel-driven model is one of the most widely used downscaling algorithms. It first establishes the relationship between low-spatial-resolution (LR) LST and kernels (e.g., reflectance, spectral indexes, albedo, emissivity, terrain factors, and others), followed by the collection of regression residuals. Then, the relationship is directly applied to high-spatial-resolution (HR) kernels based on the scale-invariant assumption, and the interpolated residuals are subsequently integrated to produce the final HR LST. The specific regression methods mainly include linear or nonlinear regression (Agam et al., 2007b; Agam et al., 2007a; Bindhu et al., 2013; Dominguez et al., 2011; Dong et al., 2020; Kustas et al., 2003), and machine learning models such as random forest (RF), support vector regression (SVR), and artificial neural network (ANN) (Dong et al., 2020; Ebrahimy and Azadbakht, 2019; Ghosh and Joshi, 2014; Hutengs and Vohland, 2016; Li et al., 2019; Yang et al., 2010; Yang et al., 2017). In recent years, various studies established local rather than global regression relationships to account for the spatial non-stationarity, either by introducing moving-windows (Chen et al., 2012; Gao et al., 2017; Jeganathan et al., 2011; Xia et al., 2019b; Zakšek and Oštir, 2012) or by using the geographically weighted regression (GWR) and its variants (Luo et al., 2021; Peng et al., 2019; Pereira et al., 2018; Wang et al., 2020; Wu et al., 2022). The kernel-driven model is simple and easy to implement. However, the scale-invariant assumption may not always hold (Chen et al., 2014; Ghosh and Joshi, 2014; Jeganathan et al., 2011; Zhou et al., 2016), and the simple interpolation of LR residuals may degrade spatial details (Agam et al., 2007b; Anderson et al., 2004; Y. Hu et al., 2023). Furthermore, the relationship requires calibration for each individual scene, potentially increasing computational burden, particularly when machine learning models are employed for regression.

The second approach is the weight-function fusion method, a spatiotemporal fusion technique that was originally developed for reflectance data. Based on the LR data at the target time and at least one LR-HR data pair at the reference time, its core idea is to predict HR values at the target time by integrating spatial discrepancies and temporal dynamics through weight functions. The spatial and temporal adaptive reflectance fusion model (STARFM) Gao et al. (2006) and its enhanced version (ESTARFM) (Zhu et al., 2010) have been introduced for LST downscaling (Li et al., 2016; Liu and Weng, 2012; Long et al., 2020; Tang et al., 2024; Wu et al., 2013; Yang et al., 2016). Recently, several studies have incorporated the intrinsic characteristics of LST to adapt these reflectance fusion models to the thermal regime (Weng et al., 2014; Wu et al., 2015; Yu et al., 2023). Compared to the kernel-driven model, the fusion method relies less on auxiliary data. However, due to frequent cloud cover and marked temporal dynamics of LST, the requirement for one or more pairs of synchronized LR and HR LST data is often difficult to fulfill, thereby constraining its applicability over large spatial extents.

The thermal mixing method assumes that the LST (or radiance) of a pixel is a linear combination of its endmembers, each weighted by the corresponding fractional area. The fractional areas of endmembers are solved from HR reflectance data and spectral indexes using the spectral unmixing method, while their LSTs (or radiances) are determined from representative pure LR pixels (Deng and Wu, 2013a; Deng and Wu, 2013b; Li et al., 2021) or the thermal disentangle approach (Mitraka et al., 2015; Pu and Bonafoni, 2021; F. Xu et al., 2024). The thermal mixing method appears robust over heterogeneous surfaces (Pu and Bonafoni, 2023), while the division of endmembers remains somewhat manual or computational demanding (F. Xu et al., 2024). Furthermore, the simple linear combination of LST, along with the assumption that the same endmember within a scene or moving window has the same LST, lacks rigorous theoretical foundation and may also introduce uncertainties.

Recently, several studies have explored physically based approaches for LST downscaling (Firozjaei et al., 2024; Y. Hu et al., 2023). For example, based on the SEB equation and Penman-Monteith equation, Y. Hu et al. (2023) derived the physical analytical expression of the differences between HR and LR LSTs, which were subsequently estimated from surface, radiation, and atmospheric parameters. The physical method has a relatively solid physical foundation and interpretability. However, it involves too many input parameters and some of which are difficult to obtain, thereby limiting its large-scale applicability and performance.

Deep learning has witnessed rapid development in recent years and has gained increasing popularity in LST downscaling. Considering the input parameters, current methods can be further divided into three groups. The first group draws inspiration from the weight-function spatiotemporal fusion method, employing different deep learning techniques to fuse multi-temporal HR and LR LSTs, including the convolutional neural network (CNN) (Yin et al., 2021), conditional variational autoencoder (Chen et al., 2022), and Swin-Unet (Hu et al., 2025). However, they also suffer from the difficulty of preparing multi-temporal paired LST images. The second group maps HR LST from the corresponding LR LST and HR auxiliary geophysical parameters (e.g., reflectance, spectral indexes, and terrain factors) using deep learning models such as the multi-layer perceptron (MLP) (Hurduc et al., 2024) and refined U-Net (Ait-Bachir et al., 2025; Li et al., 2025). Recently, by introducing the large selective kernels with dynamic modality-conditioned projections, the modality-conditioned large selective kernel network (MoCoLSK-Net) was reported to achieve the state-of-the-art (SOTA) performance for LST downscaling among deep learning models (Dai et al., 2025). The third group directly predicts HR LST from LR LST without any auxiliary parameters (Chen et al., 2024; Jahangir et al., 2025; Xu et al., 2025), thus offering greater simplicity and ease of use. However, due to the lack of auxiliary data, these algorithms essentially generate HR details from scratch, leading to generally lower performance than the other two groups.

The last one is the hybrid method, which combines the abovementioned LST downscaling methods to improve their respective performances. The prevailing approach involves combining the kernel-driven and fusion methods (Bai et al., 2015; Dong et al., 2023; Y. Li et al., 2023; Su et al., 2024; Xia et al., 2019a; Xia et al., 2018; Zhu et al., 2021). Other methods, such as integrating kernel-driven models with spatial interpolation (Hu et al., 2024) or CNN (Yu et al., 2021), combining fusion method with thermal mixing models (Quan et al., 2018; Shi et al., 2022), have also been developed. Although the hybrid approach aims to combine the advantages of different models, attention should be paid to the increased computational complexity and the potential risk of error propagation.

Although significant achievements have been made in LST downscaling, several common challenges remain. First, a benchmark dataset available to the public encompassing different scenarios with a large spatial extent for model calibration and evaluation is still unavailable, hindering fair and comprehensive comparison among models. The recent released GrokLST dataset (Dai et al., 2025) is a pioneering attempt, yet it is limited to the Heihe River Basin and the maximum upscaling factor is only 8. Table 1 summarizes the common spatial resolutions of thermal observations from sensors or satellites. To downscale LST data from medium or low spatial resolutions to high spatial resolutions of ~100 m, the upscaling factor typically needs to be 10×, 20×, or even greater. Second, the practical application of most methods across large spatial domains is hindered by high computational complexity or input parameters that are challenging to acquire. Furthermore, the performance of downscaling methods requires further improvement, especially over heterogeneous areas. Finally, most models are unable to self-assess the uncertainty in the predicted LST, which may obscure the accuracy information of the downscaled results in downstream applications.

**Table 1.** Spatial resolution (nadir) of thermal observations from representative sensors or satellites.

| High resolution (≤100 m) | Medium resolution (~1 km) | Low resolution (>1 km) |
|---|---|---|
| Landsat (100 m*) | MODIS (1 km) | GOES (2 km) |
| ASTER (90 m) | AVHRR (1 km) | Himawari (2 km) |
| ECOSTRESS (70 m) | VIIRS (750 m*) | MTG (2 km) |
|  | SLSTR (1 km) | MSG (3 km) |
|  | FY-3 (1 km) | FY-4 (4 km) |

*: The native spatial resolutions of Landsat 8/9 and VIIRS without interpolation are 100 m and 750 m, respectively.

To address the aforementioned challenges in LST downscaling, we attempted to propose a feasible solution in this study. First, three comprehensive datasets were meticulously compiled for model training and evaluation, which are expected to serve as a benchmark for future research. Subsequently, grounded in the SEB-based geophysical reasoning, we proposed the physically guided diffusion model (PGDM), a novel framework that formulates LST downscaling as an inference problem over a conditional posterior distribution. The model acts as a powerful sampler, generating HR LST under the guidance of LR LST and HR auxiliary geophysical parameters. The performance of PGDM was comprehensively evaluated and compared with several representative methods on the compiled datasets, demonstrating its effectiveness and stability. Finally, leveraging the inherent randomness of PGDM and model ensembling, we developed a scene-level uncertainty indicator to predict the downscaling errors for each individual scene.

This paper is organized as follows. Sections 2 introduces the utilized data and methods in detail. Section 3 presents the results, followed by the discussion in Section 4. Finally, Section 5 provides a brief conclusion.

## 2. Data and methods

### 2.1 SEB-based geophysical reasoning

The relationship between the HR and LR LST is modelled as follows:

$$T_s^{HR} = T_s^{LR} + \Delta T_s \tag{1}$$

where $T_s$ is LST, the superscripts "$HR$" and "$LR$" denote geophysical data at fine and coarse scales, and $\Delta T_s$ is their difference. In practice, $T_s^{LR}$ is resampled to match the spatial resolution of $T_s^{HR}$. The core principle of LST downscaling is to estimate the spatial detail $\Delta T_s$ and add it to $T_s^{LR}$ to reconstruct $T_s^{HR}$.

Based on the SEB equation and Penman-Monteith equation, and assuming that the coarse- and fine-scale pixels share the same atmospheric states and solar irradiance, $\Delta T_s$ can be physically modelled as follows (Y. Hu et al., 2023):

$$\Delta T_s = \frac{\frac{\partial T_s}{\partial R_n}\{-R_s^\downarrow dr + [R_l^\downarrow - \sigma(T_s^{LR})^4 d\varepsilon_{bb}]\} + \frac{\partial T_s}{\partial f_c}df_c + \frac{\partial T_s}{\partial r_a}dr_a + \frac{\partial T_s}{\partial r_s}dr_s}{1 + \frac{\partial T_s}{\partial R_n}4\varepsilon_{bb}\sigma(T_s^{LR})^3} \tag{2}$$

where $R_n$ is the surface net radiation, $R_s^\downarrow$ is the surface shortwave downward radiation, $r$ is the shortwave broadband albedo, $R_l^\downarrow$ is the surface longwave downward radiation, $\sigma$ is the Stefan-Boltzmann constant, $\varepsilon_{bb}$ is the surface broadband emissivity, $f_c$ presents the fractional vegetation cover, $r_a$ is the aerodynamic resistance, and $r_s$ is the surface resistance. The detailed expression of the partial derivative terms can be found in Y. Hu et al. (2023).

Although Eq. (2) provides the explicit physical expression of $\Delta T_s$, it requires numerous input parameters, many of which are difficult to obtain accurately. Consequently, both the risk of error propagation and the excessive number of input parameters hinder the practical applicability of the physical model, particularly across large spatial extents. Furthermore, the simplification during the derivation of Eq. (2), such as using LST as a substitute of aerodynamic temperature for calculating sensible heat flux, may also introduce uncertainties into the results.

According to Eq. (2) and the expression of the partial derivative terms, $\Delta T_s$ is related to $T_s^{LR}$, surface or near-surface properties (e.g., $r, \varepsilon_{bb}, f_c, r_s, r_a$) at both fine and coarse scales, atmospheric conditions, and surface radiation in theory. Among these parameters, $r$ can be estimated from surface reflectance (i.e., $\rho$) (Qu et al., 2015); $\varepsilon_{bb}$ is related to narrowband thermal emissivities (Cheng et al., 2013; Tang et al., 2011), which in turn are linked to the normalized difference vegetation index (NDVI) or $\rho$ (Cheng et al., 2018; S. Xu et al., 2024); $f_c$ is typically estimated from NDVI (Carlson and Ripley, 1997); physically modelling of both $r_s$ and $r_a$ are complex, while they are generally related to surface characteristics (e.g., $r$, NDVI, vegetation height, land cover type, etc.) and atmospheric states (e.g., wind speed, air temperature, air vapor pressure deficit, etc.) (Mu et al., 2007; Paulson, 1970; Trebs et al., 2021). Based on the above analyses, $\Delta T_s$ can be finally expressed as the function of:

$$\Delta T_s = f(T_s^{LR}, \rho^{HR}, \rho^{LR}, \text{NDXI}^{HR}, \text{NDXI}^{LR}, \text{LULC}^{HR}, \text{LULC}^{LR}, V_{ar}, V_{others}) \quad (3)$$

where $f(\cdot)$ is the nonlinear implicit function, NDXI is a group of spectral indices that reflects surface geophysical characteristics, LULC denotes the land use and land cover type, $V_{ar}$ represents the collection of atmospheric and radiation parameters, including $R_s^\downarrow$, $R_l^\downarrow$, wind speed, air temperature, air vapor pressure deficit, and others, $V_{others}$ denotes other possible parameters that are not explicitly listed here.

For multi-scale hierarchical neural networks (e.g., CNN and Swin Transformer), it is sufficient to provide only high-resolution auxiliary data, as the model can automatically extract hierarchical feature maps across different scales. Furthermore, given the generally low spatial resolution of available atmospheric products and large uncertainties of estimating radiation parameters, $V_{ar}$ was not included as inputs here; instead, $T_s^{LR}$ and HR elevation data (i.e., $H^{HR}$) were used as proxies for the atmospheric and radiation conditions. Finally, by omitting $V_{others}$, Eq. (3) is streamlined as:

$$\Delta T_s = f(T_s^{LR}, \rho^{HR}, \text{NDXI}^{HR}, H^{HR}, \text{LULC}^{HR}) \quad (4)$$

Compared to Eq. (2), the input parameters in Eq. (4) are more easily accessible, despite its implicit mathematical form. Motivated by the demonstrated success of deep learning in modeling complex implicit relationships, we adopted a data-driven strategy to estimate $\Delta T_s$ according to Eq. (4). A PGDM was employed as the core framework due to its strong generative capability, with the underlying principles elaborated in Section 2.3.

**2.2 Dataset compilation**

Three representative datasets were compiled in this study: (1) a Landsat-based dataset over mainland China in 2020 (Landsat_CN20) for model training and evaluation; (2) a Landsat-based dataset covering the selected heterogeneous regions over the globe excluding mainland China (Landsat_GLB) for model evaluation; and (3) an ASTER-based dataset covering heterogeneous regions worldwide (ASTER_GLB), designed to assess the robustness of the trained model when applied to different satellite data.

For training downscaling models to capture the implicit relationship in Eq. (4), the dataset should contain paired samples of $T_s^{HR}$, $T_s^{LR}$, $\rho^{HR}$, $\text{NDXI}^{HR}$, $H^{HR}$, and $\text{LULC}^{HR}$. To achieve this, multi-source data were used, including the Landsat 8 Level 2 Collection 2 Tier 1 (Landsat 8

L2C2T1) dataset (Malakar et al., 2018) providing $T_s^{HR}$ and $\rho^{HR}$, the Advanced Spaceborne Thermal Emission and Reflection Radiometer (ASTER) Level 2 Surface Kinetic Temperature product (AST_08 V004) (Gillespie et al., 1998) for $T_s^{HR}$, the Harmonized Landsat Sentinel-2 (HLS V002) product (Ju et al., 2025) providing $\rho^{HR}$, the Copernicus GLO-30 DEM dataset (Fahrland et al., 2022) for $H^{HR}$, and the Copernicus Global Land Service-Land Cover map (CGLS-LC) Collection 3 dataset (Buchhorn et al., 2020) for $LULC^{HR}$. Table 2 summarizes these remote sensing and auxiliary datasets, including the specific geophysical variables and their corresponding applications. All the geospatial data were resampled to 0.001° using the nearest neighbor method.

**Table 2.** Summary of the remote sensing and auxiliary datasets used in this study.

| Dataset | Variables | Usage |
|---|---|---|
| Landsat 8 L2C2T1 | SR_B2, SR_B3, SR_B4, SR_B5, SR_B6, SR_B7, ST_B10, QA_PIXEL | Compiling the Landsat_CN20 and Landsat_GLB datasets |
| AST_08 V004 | LST, QA_DataPlane | Compiling the ASTER_GLB dataset |
| HLSL30 V002 | B2, B3, B4, B5, B6, B7, Fmask | Compiling the ASTER_GLB dataset |
| HLSS30 V002 | B2, B3, B4, B8A, B11, B12, Fmask | Compiling the ASTER_GLB dataset |
| Copernicus GLO-30 DEM | DEM | Compiling the Landsat_CN20, Landsat_GLB, and ASTER_GLB datasets |
| CGLS-LC100 V3 | Discrete Classification | Compiling the Landsat_CN20 dataset |

In this study, $\rho^{HR}$ was determined from the band configurations of Landsat 8 and HLS data, including surface reflectance at the blue, green, red, near infrared (NIR), 1.6 μm shortwave infrared (SWIR1), and 2.2 μm shortwave infrared (SWIR2) bands. Subsequently, the spectral indexes, including NDVI, the normalized difference water index (NDWI), and the normalized difference moisture index (NDMI), were calculated from $\rho^{HR}$ to construct $NDXI^{HR}$. According to Table 1, two scaling factors (i.e., 10× and 20×) were chosen in this study to better reflect the resolution differences among sensors. Then, $T_s^{LR}$ was obtained from $T_s^{HR}$ based on the energy conservation principle:

$$T_{s,ij}^{LR} = \sqrt[4]{\frac{\sum_{x=1}^{X}\sum_{y=1}^{Y}\varepsilon_{bb,xy}^{HR}\left(T_{s,xy}^{HR}\right)^4}{\sum_{x=1}^{X}\sum_{y=1}^{Y}\varepsilon_{bb,xy}^{HR}}} \approx \sqrt[4]{\frac{\sum_{x=1}^{X}\sum_{y=1}^{Y}\left(T_{s,xy}^{HR}\right)^4}{XY}} \quad (5)$$

where the subscripts $i$ and $j$ denote the location of a coarse pixel in the original $T_s^{LR}$ image without resampling, a coarse pixel is assumed to comprise $X \times Y$ fine pixels, and the subscripts $x$ and $y$ denote relative positions of the constituent fine pixels in the coarse pixel. Emissivity is omitted for simplicity during the simulation of $T_s^{LR}$, and the corresponding accuracy loss will be further discussed in Section 4.1.

The first dataset was Landsat_CN20 covering mainland China for model training and evaluation, as the diverse land cover types and substantial elevation differences across this region make it an ideal case for conducting LST downscaling experiments. Based on Landsat 8 L2C2T1 images acquired in 2020, together with the corresponding DEM and LULC data, a total of 22,909 clear-sky image patches of 160×160 HR pixels were compiled. The spatial and temporal balance of image patch distribution was carefully considered, with detailed methods given in Supplementary Text S1. Fig. 1(a) shows the spatial distribution of involved Landsat 8 images, in which most of mainland China is covered except for certain southern regions where Landsat imagery is unavailable due to frequent cloud coverage. Fig. 1(b) compares the number of images and patches in each month. Although the availability of images varies considerably by month, the final number of patches is relatively balanced. Fewer patches are available in January and December compared to other months, which is acceptable given that LST heterogeneity is generally lower in winter. Fig. S1 provides an example data pair in the Landsat_CN20 dataset. The Landsat_CN20 dataset was further randomly split into training, validation, and test sets with a ratio of 70%:15%:15%.

The Landsat_GLB dataset was constructed from the same data sources as Landsat_CN20, but over 20 heterogeneous regions worldwide outside China to evaluate the model's performance and robustness. Different from the Landsat_CN20 dataset, the size of HR data in Landsat_GLB was 480×480 pixels, with the corresponding LR LST data sized at 48×48 and 24×24 pixels for 10× and 20× downscaling, respectively. Apart from the first two Landsat-based datasets, the ASTER_GLB dataset was constructed based on the ASTER LST and HLS surface reflectance data to simulate practical application scenarios, which encompassed 20 additional heterogeneous regions worldwide. The methods for compiling these two datasets are also provided in Supplementary Text S1. Fig. 2 shows the spatial distribution of all 40 evaluation regions in both the Landsat_GLB and ASTER_GLB datasets, with the corresponding basic information given in Supplementary Tables S1 and S2. The relatively balanced spatial distribution of these regions, together with their diverse land cover types, climatic conditions, and topographic characteristics, is expected to provide an objective and comprehensive evaluation.

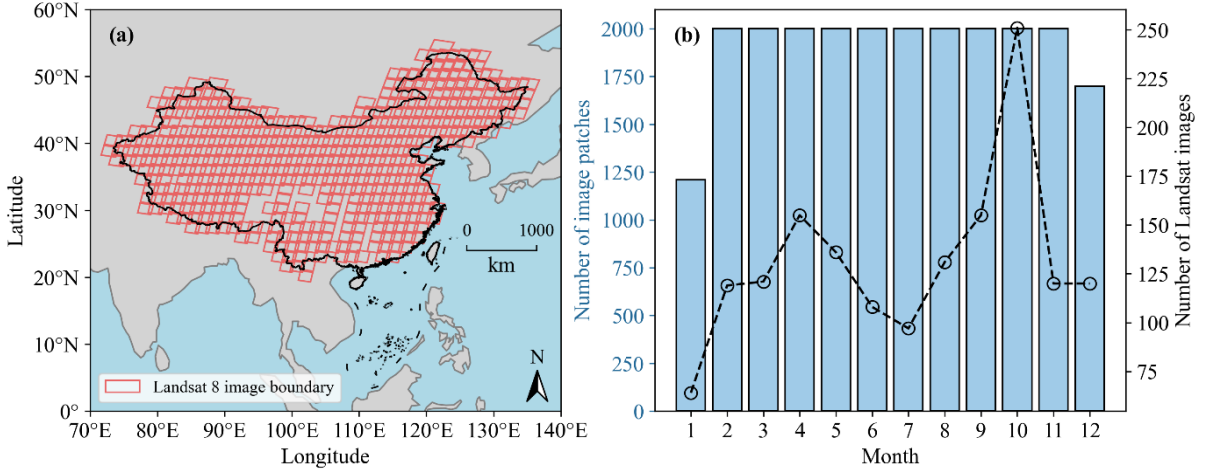

**Fig. 1.** (a) Spatial and (b) temporal distribution of the compiled Landsat_CN20 dataset.

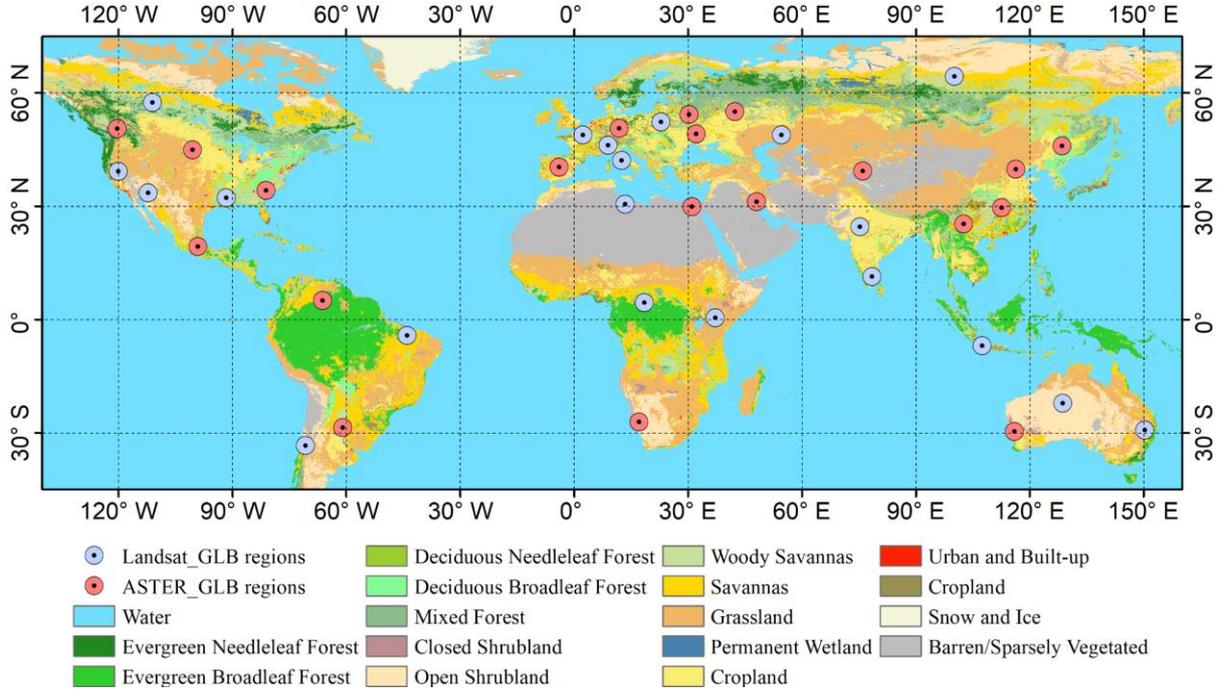

**Fig. 2.** Spatial distribution of the evaluation regions in the Landsat_GLB and ASTER_GLB datasets.

## 2.3 The physically guided diffusion model

Hereafter, images are denoted in bold for clarity. LST downscaling is inherently an ill-posed inverse problem, as a single LR LST observation corresponds to infinitely many potential HR thermal fields. Based on the geophysical reasoning results in Eqs. (1) and (4), the LST downscaling task can be reformulated as estimating the conditional posterior distribution $p(\boldsymbol{T}_s^{HR}|\boldsymbol{T}_s^{LR}, \boldsymbol{\mathcal{G}}^{HR})$, where $\boldsymbol{\mathcal{G}}^{HR}$ denotes a collection of HR geophysical guidance variables (i.e., $\boldsymbol{\rho}^{HR}$, $\mathbf{NDXI}^{HR}$, $\boldsymbol{H}^{HR}$, $\mathbf{LULC}^{HR}$). In this study, we employ the residual shifting diffusion model (ResShift) (Yue et al., 2025; Yue et al., 2023) to sample from this high-dimensional posterior distribution, thereby generating high-quality downscaling results by leveraging both the observational constraints from $T_s^{LR}$ and the geophysical knowledge embedded in $\mathcal{G}^{HR}$.

### 2.3.1 Diffusion process

The diffusion model typically comprises a forward Markov process that gradually perturbs the target image by adding Gaussian noise, and a corresponding reverse process that iteratively denoises the image to reconstruct its original state (Ho et al., 2020). Through this iterative destruction-reconstruction process, the model learns to invert the noise injection process, thereby enabling efficient sampling from complex, high-dimensional data distributions.

In the forward process, the initial $T_s^{HR}$ is gradually transformed into the final state, which is a prior distribution related to $T_s^{LR}$, by progressively shifting their residuals. Corresponding, the forward Markov chain with length $T$ is defined as follows:

$$q(T_{s,t}^{HR}|T_{s,t-1}^{HR}, T_{s,0}^{HR}, T_s^{LR}) = \mathcal{N}(T_{s,t}^{HR}; T_{s,t-1}^{HR} - \alpha_t \Delta T_s, \kappa^2 \alpha_t I) \tag{6}$$

where the timestep $t$ belongs to $[1, 2, …, T]$, $T_{s,0}^{HR}$ is the initial state of HR LST, $\mathcal{N}(x; \mu, \sigma)$ denotes the normal distribution for the variable $x$ with the mean value of $\mu$ and standard deviation (STD) value of $\sigma$, $\alpha_t$ and $\kappa$ are hyperparameters that controls the shifting speed and noise intensity during the transition process, $I$ is the identity matrix. As the Markov chain is integrable, the direct transition between $T_{s,t}^{HR}$ and $T_{s,0}^{HR}$ can be straightforwardly expressed as follows:

$$q(T_{s,t}^{HR}|T_{s,0}^{HR}, T_s^{LR}) = \mathcal{N}(T_{s,t}^{HR}; T_{s,0}^{HR} - \eta_t \Delta T_s, \kappa^2 \eta_t I) \tag{7}$$

where $\eta_t$ is a derived hyperparameter and $\eta_t - \eta_{t-1} = \alpha_t$. To ensure a smooth transition from $T_{s,0}^{HR}$ to $T_{s,T}^{HR}$, $\eta_1$ and $\eta_T$ are constrained to approach 0 and 1, respectively.

The reverse denoising process starts from $T_{s,T}^{HR}$ and recovers $T_{s,0}^{HR}$ through iterative refinements. According to Bayes' theorem, the reverse Markov process can be expressed as:

$$p(T_{s,t-1}^{HR}|T_{s,t}^{HR}, T_{s,0}^{HR}, T_s^{LR}) = q(T_s^{HR}|T_{s,t-1}^{HR}, T_{s,0}^{HR}, T_s^{LR}) \frac{q(T_{s,t-1}^{HR}|T_{s,0}^{HR}, T_s^{LR})}{q(T_s^{HR}|T_{s,0}^{HR}, T_s^{LR})} \tag{8}$$

By substituting Eqs. (6) and (7) into Eq. (8), the specific expression of $p(T_{s,t-1}^{HR}|T_{s,t}^{HR}, T_{s,0}^{HR}, T_s^{LR})$ is derived as follows:

$$p(T_{s,t-1}^{HR}|T_{s,t}^{HR}, T_{s,0}^{HR}, T_s^{LR}) = \mathcal{N}\left(T_{s,t-1}^{HR}; \frac{\eta_{t-1}}{\eta_t} T_{s,t}^{HR} + \frac{\alpha_t}{\eta_t} T_{s,0}^{HR}, \frac{\kappa^2 \alpha_t \eta_{t-1}}{\eta_t} I\right) \tag{9}$$

Regarding this normal distribution, the mean value $\mu_t = (\eta_{t-1} T_{s,t}^{HR} + \alpha_t T_{s,0}^{HR})/\eta_t$, and the STD value $\sigma_t = \kappa \sqrt{\alpha_t \eta_{t-1}}/\sqrt{\eta_t}$. As $\sigma_t$ is readily available once the hyperparameters are specified, thus the key of solving $p(T^{HR}_{s,t-1}|T^{HR}_{s,t}, T^{HR}_{s,0}, T^{LR}_s)$ lies in the estimation of $\mu_t$. Following the recommendation from previous studies (Yue et al., 2025; Yue et al., 2023) as well as the SEB-based geophysical reasoning results in Section 2.1, $\mu_t$ is reparametrized as:

$$\mu_\theta(T^{HR}_{s,t}, T^{LR}_s, \mathcal{G}^{HR}, t) = \frac{\eta_{t-1}}{\eta_t} T^{HR}_{s,t} + \frac{\alpha_t}{\eta_t} f_\theta(T^{HR}_{s,t}, T^{LR}_s, \mathcal{G}^{HR}, t) \tag{10}$$

where $f_\theta$ represents the learnable denoising neural network.

The complete reverse process in ResShift initiates from $T^{HR}_{s,T}$ that is closely related to $T^{LR}_s$, and then iteratively estimates the state of each previous timestep through Eqs. (9) and (10) with $f_\theta$, finally recovering $T^{HR}_{s,0}$. Compared to traditional diffusion models whose reverse process starts from pure Gaussian noise (Ho et al., 2020; Li et al., 2022; Saharia et al., 2022; Song et al., 2020), the modification in ResShift is particularly reasonable for image super-resolution tasks and effectively shortens the length of the Markov chain. Fig. 3 provides a simplified schematic illustration of the forward and reverse processes as defined in ResShift.

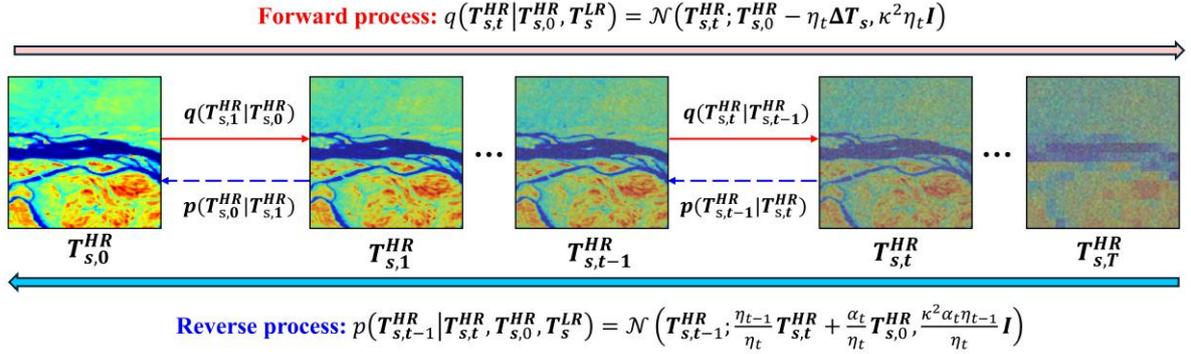

**Fig. 3.** The forward and reverse diffusion processes. The forward process is to gradually perturb the target image (i.e., $T^{HR}_{s,0}$) by adding Gaussian noise and shifting residuals. The reverse process is to iteratively remove the random noise and recover the initial target image.

### 2.3.2 The denoising network

In this study, $f_\theta$ is designed as a dual-branch encoder-decoder architecture for multi-modal fusion (Fig. 4). Specifically, the encoder comprises a state-aware branch and a geophysical-prior branch, respectively extracting multiple-scale feature maps from the dynamic temperature state (i.e., $T^{HR}_{s,t}$ and timestep $t$) and the static geophysical conditional data (i.e., $T^{LR}_s$ and $\mathcal{G}^{HR}$). The decoder is then used to transform the encoded futures back into the target output space for estimating $T^{HR}_{s,0}$.

In the geophysical-prior encoder, the categorical **LULC**$^{HR}$ in $\mathcal{G}^{HR}$ is first embedded to continuous values using the Embedding layer in Pytorch. Subsequently, $\mathcal{G}^{HR}$ is concatenated with the upsampled $T_s^{LR}$ to form the inputs advised by Eq. (4). During training, the channel-wise dropout strategy (Guo et al., 2024) is adopted for $\mathcal{G}^{HR}$, serving as a regularization method and enabling the network to flexibly handle varying sets of $\mathcal{G}^{HR}$. After the preprocessing, a convolutional stem (Conv Stem) is used to extract shallow features from the inputs. The stem consists of two convolutional layers with an activation layer in between, which unify the channel numbers of shallow features to $C_{base}$. The extracted features are then passed through sequential residual blocks (ResBlock) which double the channel number, and downsampling convolutional layers with a stride of 2 (Conv Down) that reduce the spatial dimensions of feature maps by half. As shown in Fig. 5, each ResBlock consists of two sequential "Normalization-Activation-Convolution Layer" units, where the group normalization and the sigmoid linear unit (SiLU) activation function are used. At the bottom of the encoder, the Multi-head Non-local (MHNL) module is introduced to capture long-range dependencies and therefore compensate for the locality of convolutional networks, which adapts the original Non-local block (Wang et al., 2018) by introducing multiple parallel heads (Dosovitskiy et al., 2020; Vaswani et al., 2017).

The state-aware encoder shares a similar architecture with the geophysical-prior encoder, while the original ResBlock is replaced by the time-adaptive ResBlock (TA-ResBlock) to inform the dynamic state-aware branch about the expected noise level of $T_{s,t}^{HR}$. Specifically, $t$ is first embedded using the sinusoidal positional encoding (Vaswani et al., 2017), followed by a globally shared MLP. The embedded $t$ then passes through the time adaptive module, which adopts an adaptive group norm block with gating mechanism (Tian et al., 2025) to inject the conditional timestep information into the network.

The features from these two encode branches are fused by direct addition across multiple scales, which are then concatenated with the upsampled features achieved by a "Convolution Layer-Pixel Shuffle" unit (Shi et al., 2016) at corresponding scales in the decoder. The skip connection strategy follows the standard U-Net decoding process (Ronneberger et al., 2015), enabling the progressive recovery of spatial resolution while preserving semantic information from deeper layers. The final projection head is a simple 1×1 convolutional layer. The direct residual connection between the upsampled $T_s^{LR}$ and outputs after the projection head ensures that the model learns to predict $\Delta T_s$ rather than $T_{s,0}^{HR}$, thereby following and paradigm in Eq. (4) and facilitating faster convergence during training.

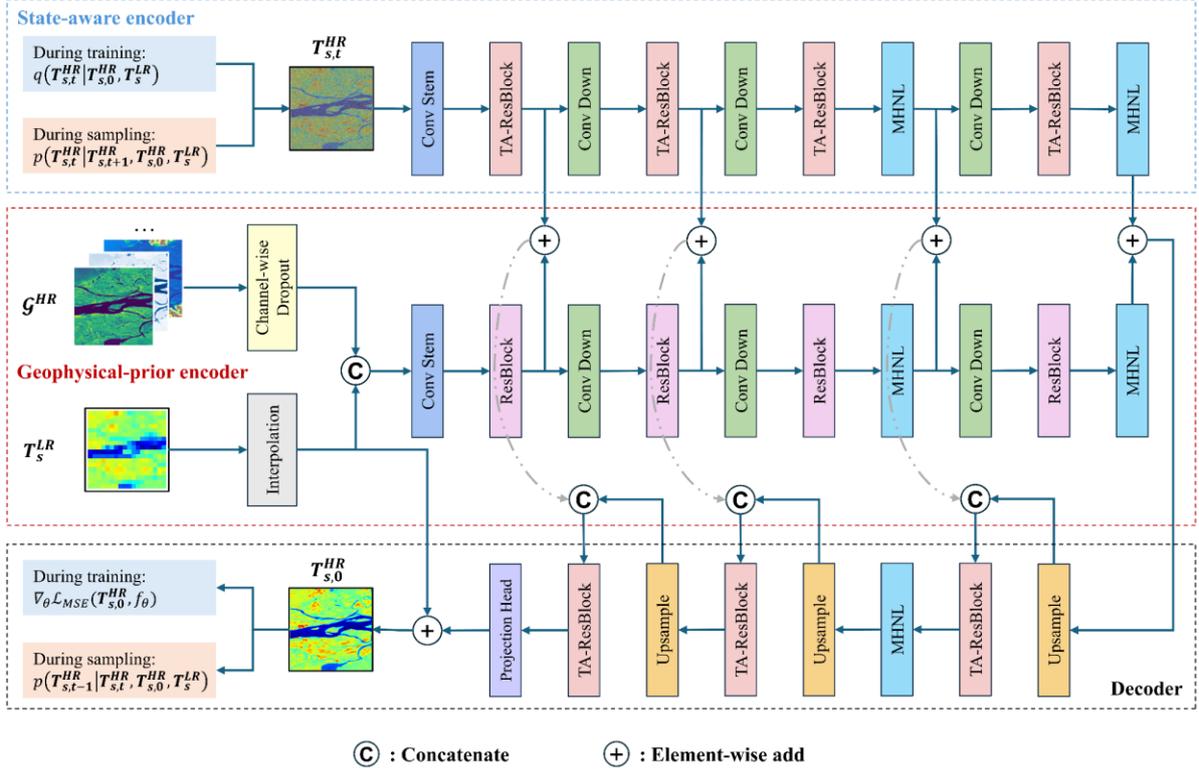

**Fig. 4.** Structure of the denoising neural network.

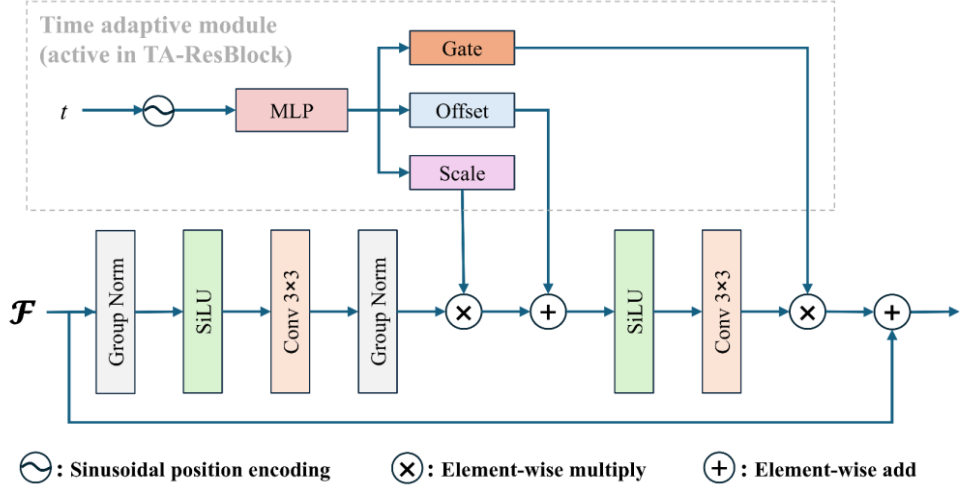

**Fig. 5.** Structure of the ResBlock and TA-ResBlock. $\mathcal{F}$ denotes the input feature maps.

### 2.3.3 Training and hyperparameter configurations

The denoising neural network $f_\theta$ was trained on the Landsat_CN20 train set. $T_{s,t}^{HR}$ was first generated from random $t$ and Gaussian noise using Eq. (7), and $f_\theta$ was then trained to predict $T_{s,0}^{HR}$ from $T_{s,t}^{HR}$, $T_s^{LR}$, $\mathcal{G}^{HR}$, and $t$. The bilinear interpolation method was chosen to upsample $T_s^{LR}$ as it provides good spatial continuity. The loss function was simply chosen as the mean squared error (MSE) as recommended by previous studies (Ho et al., 2020; Yue et al., 2023). The model parameters were updated using the AdamW optimizer for 300 epochs. The initial learning rate was set to $10^{-4}$ and updated using a cosine annealing scheduler. $C_{base}$ in $f_\theta$ was set to 32. During training, the random drop rate of individual channels in $\mathcal{G}^{HR}$ was set to 20%.

After each epoch's training, the initial state $T_{s,0}^{HR}$ was generated by iteratively applying Eq. (9), and the corresponding MSE loss was tracked on the validation set. The early stopping strategy was applied when the validation loss did not decrease for 20 consecutive epochs to avoid overfitting. Due to the introduction of $\mathcal{G}^{HR}$, the total timesteps of the forward and reverse processes were empirically reduced to 3 in this work, which further improved the computational efficiency during model inference. In contrast to natural images, the numerical variation in LST image patches is considerably more limited in general. To ensure a smooth transition between HR and LR LST in the Markov chain, $\kappa$ was set to 0.02 to reduce the noise intensity, $\sqrt{\eta_t}$ was set to [0.05, 0.60, 0.99]. All computations in this study were performed on a device equipped with an Intel Core i9-10980XE CPU @ 3.00 GHz, 256 GB RAM, and an NVIDIA GeForce RTX 4090 GPU.

## 2.4 Evaluation and comparison

The performance of PGDM was evaluated on the three compiled datasets using the widely adopted "upscaling-downscaling" strategy (Agam et al., 2007a; Dai et al., 2025; Dominguez et al., 2011; Dong et al., 2023; Dong et al., 2020; Hu et al., 2024; Kustas et al., 2003; Xia et al., 2019a). Specifically, $T_s^{LR}$ was first simulated from $T_s^{HR}$ based on Eq. (5). The upscaled image ($T_s^{LR}$) was then downscaled back to the fine scale and assessed against the original $T_s^{HR}$. Meanwhile, several representative LST downscaling methods were also reproduced in this study for comparative analysis, including the bilinear interpolation method, the kernel-driven model, the dual-layer composite framework (DCF) method that combines the bilinear interpolation and kernel-driven methods to account for both spatial heterogeneity and autocorrelation (Hu et al., 2024), and the SOTA deep learning model MoCoLSK-Net (Dai et al., 2025). The core idea of the kernel-driven model and MoCoLSK-Net is to add spatial details derived from auxiliary data to $T_s^{LR}$. Consequently, these selected methods (except for bilinear interpolation) follow the paradigms given in Eqs. (1) and (4), making them more comparable. The other LST downscaling methods (e.g., the weight-function fusion method, the thermal unmixing method, and the physical method) and evaluation strategies (e.g., T-based validation) are not included in this study, and the reason will be further discussed in Section 4.3.

For the kernel-driven model, a linear regression model was employed to establish the relationship between $T_s^{LR}$ and LR kernels on the Landsat_CN20 dataset, given the relatively small size of the LR data (i.e., 16×16 pixels for 10× super-resolution and 8×8 pixels for 20× super-resolution) and the fact that the accuracy gap between linear regression and machine learning models is not significant (Dong et al., 2020). The kernel-driven models in this case are more comparable to local models with medium or large window sizes, which are generally expected to achieve better downscaling results over heterogeneity surfaces than global models as they account for the spatial non-stationarity (Duan and Li, 2016; Gao et al., 2017; Jeganathan et al., 2011; Zhang et al., 2020). For the other two datasets with larger spatial extents, both the linear regression and RF models were adopted. The residuals at coarse scales in kernel-driven models were resampled to fine scales using the bilinear interpolation method, as previous studies reported that different interpolation methods (e.g., bilinear, spline, and Kriging) generally introduced only minimal differences (Duan and Li, 2016; Tang et al., 2021).

To reproduce the DCF method, the downscaled LST from the kernel-driven model was further combined with the bilinearly interpolated LST through weighted averaging, thereby accounting for both spatial heterogeneity and spatial autocorrelation. Given that the weighting coefficients of the DCF method were only calibrated over a small region (i.e., Beijing) in the original study (Hu et al., 2024), we recalibrated them on the comprehensive Landsat_CN20 training set to improve model's performance.

The training procedure of MoCoLSK-Net was kept consistent with the denoising network, using the same number of epochs, initial learning rate, scheduler, optimizer, and early stopping strategy. In the original work (Dai et al., 2025), the maximum scaling factor was 8×. However, we found that for 10× and 20× super-resolution, retaining the original network structure would result in an excessively large number of parameters and floating-point operations (FLOPs), creating computational overheads that hinder efficient deployment on our devices. To facilitate training and ensure a relatively fair comparison across models, both the number of stages and the number of residual blocks in each residual group of MoCoLSK-Net were reduced to 2 in this study. Table 3 summarizes the model parameters and FLOPs of MoCoLSK-Net and PGDM, showing that MoCoLSK-Net remains more computationally demanding even after slimming, while PGDM is more lightweight.

**Table 3.** Comparison of model parameters and FLOPs of MoCoLSK-Net and PGDM. In MoCoLSK-Net, certain modules such as the up-projection layer cause the computational complexity to grow with super-resolution factors. The FLOPs of PGDM are calculated over all timesteps.

| Model | Parameters | FLOPs |
| --- | --- | --- |
| MoCoLSK-Net | 15.61M (10×) and 41.29M (20×) | 218.12G (10×) and 556.25 (20×) |
| PGDM | 12.17M | 57.86G |

The evaluation metrics included the root mean squared error (RMSE), bias, structural similarity index (SSIM), and the proposed physical consistency loss ($\mathcal{L}_{phy}$). RMSE and bias were used to assess the total uncertainty and accuracy of the downscaling results at the pixel scale, while SSIM was employed to evaluate the spatial structural similarity at the image scale. To examine the physical consistency during the LST downscaling process, the downscaled LST was further reaggregated into coarse scales using Eq. (5) and then compared with the original $T_s^{LR}$, resulting in the physical consistency loss:

$$\mathcal{L}_{phy} = \sqrt{\frac{1}{N \times M} \sum_{k=1}^{N} \left\| \mathcal{D}(\widehat{T}_{s,k}^{HR}) - T_{s,k}^{LR} \right\|_2^2} \tag{11}$$

where $N$ is the number of images in the dataset, $M$ is the number of pixels in LR images, the subscripts $k$ denotes the image index, $\mathcal{D}(\cdot)$ denotes the degradation operator of LST as shown in Eq. (5), $\widehat{T}_s^{HR}$ is the HR LST predicted from the downscaling models. Although rarely addressed in previous studies, the metric $\mathcal{L}_{phy}$ is particularly informative, as it quantifies the extent to which a model complies with the original LR LST in the downscaling process.

## 3. Results

### 3.1 Performances on the Landsat_CN20 test set

The channel-wise dropout of $\mathcal{G}^{HR}$ during model training allows for the assessment of the relative importance of HR auxiliary parameters. Based on the Landsat_CN20 test set, Table 4 provides the accuracy metrics of PGDM under different input combinations for $\mathcal{G}^{HR}$. As both bias and $\mathcal{L}_{phy}$ are very close to zero for different schemes, only RMSE and SSIM are included for simplicity. For each scheme, the downscaling error at 20× is larger than that at 10×, which is reasonable as coarser resolutions entail greater loss of information, thereby increasing the challenge of downscaling. The relative importance of HR auxiliary parameters can be ranked as $\rho^{HR}$ > $\mathbf{NDXI}^{HR}$ > $H^{HR}$ > $\mathbf{LULC}^{HR}$. $\rho^{HR}$ contains detailed surface biophysical properties, thus playing a key role in LST downscaling. Although $\mathbf{NDXI}^{HR}$ is also a good biophysical indicator, only three indices are used here and provide less information than $\rho^{HR}$. $H^{HR}$ is also beneficial for LST downscaling, as it affects the atmospheric states such as air temperature. Finally, $\mathbf{LULC}^{HR}$ appears to be less informative, as it does not provide specific surface properties and remains constant throughout the year. Furthermore, adding $\mathbf{LULC}^{HR}$ to other HR parameters does not provide evident gains when comparing Schemes 1 and 5, or Schemes 8 and 9. The combination of $H^{HR}$ with either $\mathbf{NDXI}^{HR}$ (Scheme 6) or $\rho^{HR}$ (Scheme 7) outperforms the use of a single parameter, as they affect LST distribution from different perspectives. Compared to Scheme 7, the addition of $\mathbf{NDXI}^{HR}$ in Scheme 8 only achieves minimal accuracy boosting. A possible explanation is that $\mathbf{NDXI}^{HR}$ is derived from $\rho^{HR}$ through simple nonlinear functions, a relationship that can be easily captured by deep neural network. Therefore, for deep learning models capable of automatic feature extraction, $\mathbf{NDXI}^{HR}$ offers little additional information beyond what is already contained in $\rho^{HR}$. Based on the above analyses, Schemes 8 are recommended as the preferred choices for practical applications.

**Table 4.** Performances of PGDM under different input combinations for $\mathcal{G}^{HR}$.

| Scheme | Input combination | For 10× / 20× downscaling | |
|---|---|---|---|
| | | RMSE (K) | SSIM |
| 1 | $H^{HR}$ | 1.069 / 1.402 | 0.929 / 0.903 |
| 2 | $\mathbf{LULC}^{HR}$ | 1.323 / 1.600 | 0.895 / 0.878 |
| 3 | $\mathbf{NDXI}^{HR}$ | 1.017 / 1.300 | 0.936 / 0.918 |
| 4 | $\rho^{HR}$ | 0.699 / 0.874 | 0.966 / 0.958 |
| 5 | $H^{HR}$, $\mathbf{LULC}^{HR}$ | 1.031 / 1.321 | 0.932 / 0.909 |
| 6 | $H^{HR}$, $\mathbf{NDXI}^{HR}$ | 0.758 / 1.026 | 0.962 / 0.943 |
| 7 | $H^{HR}$, $\rho^{HR}$ | 0.620 / 0.763 | 0.973 / 0.966 |
| 8 | $H^{HR}$, $\mathbf{NDXI}^{HR}$, $\rho^{HR}$ | 0.610 / 0.749 | 0.973 / 0.967 |
| 9 | $H^{HR}$, $\mathbf{NDXI}^{HR}$, $\rho^{HR}$, $\mathbf{LULC}^{HR}$ | 0.607 / 0.742 | 0.974 / 0.968 |

Subsequently, performances of different LST downscaling methods were compared on the Landsat_CN20 test set, with the corresponding accuracy metrics summarized in Table 5. Given the relatively limited contribution of $\mathbf{LULC}^{HR}$ and the difficulty of linear regression models in handling categorical variables, the HR auxiliary parameters used in all downscaling methods except bilinear interpolation include $\boldsymbol{\rho}^{HR}$, $\mathbf{NDXI}^{HR}$, and $\boldsymbol{H}^{HR}$. As the biases of all methods are close to zero, the following analyses focus on the other three metrics. The simple bilinear interpolation yields the lowest downscaling accuracy for both 10× and 20× cases, whereas the kernel-driven model performs slightly better by incorporating HR kernel information. The DCF method outperforms the two aforementioned approaches by jointly considering spatial autocorrelation and heterogeneity. For 10× and 20× downscaling, the optimal weighting parameters of the kernel-driven model in DCF are 0.57 and 0.50, respectively. However, all three methods fail to strictly maintain physical consistency during downscaling, as indicated by their relatively higher $\mathcal{L}_{phy}$ values. In comparison, the two deep learning models generally achieve satisfactory downscaling results with better accuracy metrics, highlighting their strong capability to effectively capture the complex implicit relationship described in Eq. (4). PGDM performs best among all methods with the lowest RMSE (0.610 K for 10× and 0.749 K for 20×), highest SSIM (0.973 for 10× and 0.967 for 20×), and lowest $\mathcal{L}_{phy}$ (0.025 K for 10× and 0.042 K for 20×). The MoCoLSK-Net also shows promising downscaling performance, but its RMSE for 20× downscaling decreases relatively quickly compared to PGDM, which may be attributed to the difficulty of its up-projection and down-projection layers in preserving accuracy at extremely larger scaling factors. Consequently, PGDM achieves slightly better downscaling performance than the SOTA deep learning model while maintaining a lightweight architecture.

**Table 5.** Performances of different LST downscaling methods over the Landsat_CN20 test set.

| Method | For 10× / 20× downscaling | | | |
| --- | --- | --- | --- | --- |
| | RMSE (K) | Bias (K) | SSIM | $\mathcal{L}_{phy}$ (K) |
| Bilinear interpolation | 1.424 / 1.691 | 0.011 / 0.015 | 0.885 / 0.874 | 0.384 / 0.368 |
| Kernel-driven (Linear) | 1.237 / 1.633 | 0.011 / 0.015 | 0.915 / 0.884 | 0.215 / 0.195 |
| DCF (Linear) | 0.987 / 1.252 | 0.011 / 0.015 | 0.941 / 0.925 | 0.238 / 0.234 |
| MoCoLSK-Net | 0.633 / 0.897 | -0.004 / -0.009 | 0.971 / 0.954 | 0.035 / 0.047 |
| PGDM (Proposed) | 0.610 / 0.749 | 0.000 / 0.007 | 0.973 / 0.967 | 0.025 / 0.042 |

For a more detailed comparison among different methods, Fig. 6 provides the RMSE-based ranking of the selected downscaling methods across all 3437 scenes in the Landsat_CN20 test set. PGDM achieves the lowest RMSE in most scenes, with 2027 and 3225 cases for 10× and 20× downscaling, respectively. The greater advantage of PGDM at 20× downscaling suggests its relative stability in handling the more challenging downscaling task. MoCoLSK-Net and the DCF method generally yield the second- and third-lowest RMSEs, whereas the performances of the kernel-driven model and bilinear interpolation are less favorable. Fig. 7 further illustrates the distribution of RMSE values for each method across all scenes. Consistent with the patterns observed in Fig. 6, PGDM exhibits the most stable performance with low RMSE. MoCoLSK-Net also performs well, except for a few relatively large outliers. For the other three methods, both the magnitude and variability of RMSE values are greater.

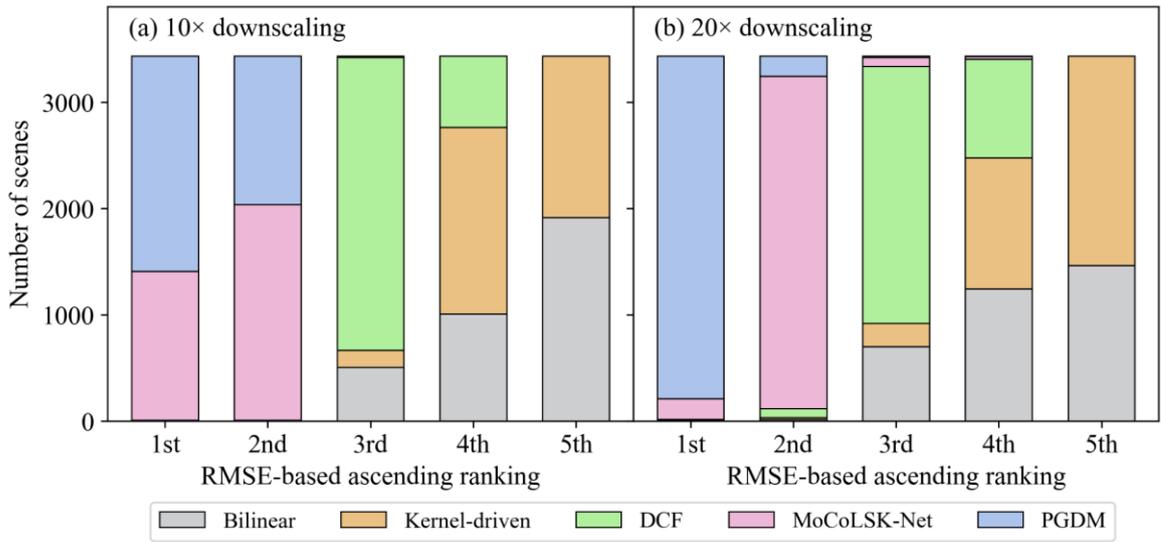

**Fig. 6.** RMSE-based ascending ranking of different downscaling methods for all scenes in the Landsat_CN20 test set under (a) 10× and (b) 20× downscaling.

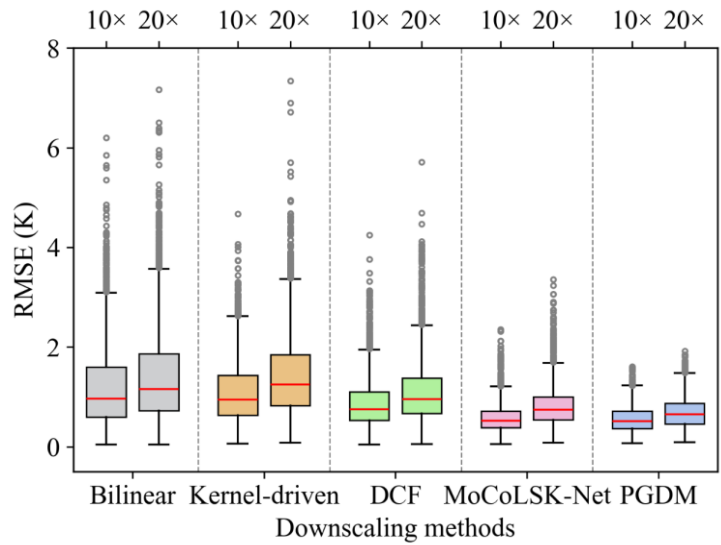

**Fig. 7.** Distribution of RMSE values for different downscaling methods across all scenes in the Landsat_CN20 test set.

## 3.2 Performances on the Landsat_GLB dataset

The performances of different models were then compared on the Landsat_GLB dataset to evaluate how deep learning models behave under out-of-distribution (OOD) conditions. Given the larger sizes of LR data in Landsat_GLB (i.e., 48×48 for 10× super-resolution and 24×24 for 20× super-resolution), both the linear regression and RF algorithms were employed in the kernel-driven model. Accordingly, the downscaling results of the kernel-driven and DCF methods were categorized into linear and RF variants. The hyperparameters of RF followed the default settings from the scikit-learn package without further optimization based on two considerations. First, our experiments indicated that the hyperparameter configuration of the RF model generally had relatively little impact on the final accuracy. Second, hyperparameter tuning via grid search or Bayesian optimization considerably reduces computational efficiency during the inference stage, as the kernel-driven model requires retraining a separate model for each scene.

The evaluation results of these seven methods are provided in Table 6. Compared to Table 5, the accuracy metrics decrease in all cases, mainly due to the larger spatial coverage and generally higher heterogeneity of the study regions. Nevertheless, the performance ranking remains consistent, following the order of PGDM, MoCoLSK-Net, DCF, kernel-driven model, and bilinear interpolation. PGDM achieves the lowest RMSE (0.910 K for 10× and 1.126 K for 20×) and highest SSIM (0.949 for 10× and 0.937 for 20×), and its $\mathcal{L}_{phy}$ is also close to zero. Following PGDM, MoCoLSK-Net also demonstrated promising downscaling performance. The remaining methods generally result in higher RMSE and lower SSIM, accompanied by reduced physical consistency in the downscaling process. For both the kernel-driven and DCF models, the accuracy discrepancy between the linear and nonlinear variants is not significant, which is consistent with previous studies (Dong et al., 2020; Hutengs and Vohland, 2016).

**Table 6.** Performances of different LST downscaling methods over the Landsat_GLB dataset.

| Method | For 10× / 20× downscaling | | | |
| --- | --- | --- | --- | --- |
| | RMSE (K) | Bias (K) | SSIM | $\mathcal{L}_{phy}$ (K) |
| Bilinear interpolation | 2.096 / 2.579 | 0.023 / 0.034 | 0.793 / 0.768 | 0.636 / 0.669 |
| Kernel-driven (Linear) | 1.902 / 2.151 | 0.023 / 0.034 | 0.848 / 0.831 | 0.320 / 0.300 |
| Kernel-driven (RF) | 1.850 / 2.020 | -0.119 / -0.091 | 0.864 / 0.854 | 0.809 / 0.837 |
| DCF (Linear) | 1.421 / 1.702 | 0.023 / 0.034 | 0.899 / 0.888 | 0.360 / 0.393 |
| DCF (RF) | 1.445 / 1.800 | -0.058 / -0.028 | 0.904 / 0.886 | 0.598 / 0.624 |
| MoCoLSK-Net | 1.026 / 1.353 | -0.002 / -0.003 | 0.939 / 0.915 | 0.060 / 0.073 |
| PGDM (Ours) | 0.910 / 1.126 | 0.004 / 0.017 | 0.949 / 0.937 | 0.041 / 0.079 |

Fig. 8 shows the 10× downscaling results of all seven methods over Santiago, Chile, a highly heterogeneous region encompassing urban areas, savannas, and grasslands, with the Andes Mountains rising along its eastern boundary. The simplest bilinear interpolation method produces overly smooth downscaling results that lack detailed features, with large discrepancies especially over heterogeneous regions. It also exhibits reduced thermal contrast compared to the ground truth, that is, overestimating low LST values and underestimating high LST values. The kernel-driven model based on linear regression yields some outliers (or granular patterns) in downscaling results, with extremely low values around 160 K (not shown in the figure) and high values above 350 K. Such misestimations or noise of linear models have also been observed by previous studies (Ghosh and Joshi, 2014; Hu et al., 2024), indicating the instability of this model. Consequently, the linear DCF model exhibits similar patterns. The RF-based kernel-driven and DCF models generally avoid such outliers; however, they also exhibit reduced thermal contrast compared to the ground truth. A possible explanation is that the machine learning models in kernel-driven methods are trained using coarse-scale data with inherently lower thermal contrast. As a result, the models generally fail to fully recover the larger thermal contrast when applied at the finer scale (Hutengs and Vohland, 2016; Yoo et al., 2020). Furthermore, the RF-based methods generally underestimate LST in this case, leading to negative biases. In contrast, the two deep learning models perform more stably, with more homogeneous error maps and most data points clustering around the 1:1 line in the scatter plot. PGDM achieves the lowest RMSE with 1.36 K and highest SSIM with 0.90, which effectively recovers thermal details with less uncertainty. The RMSE and SSIM of MoCoLSK-Net are 1.59K and 0.88, respectively.

Fig. 9 further compares the 20× downscaling results of all methods over Lake Tahoe, America, which is also characterized by various land cover types (i.e., grasslands, savannas, and water bodies) with complex terrains. All methods except bilinear interpolation recover the spatial details of LST. However, as shown in the error maps, all methods except PGDM shows large downscaling uncertainty especially at the boundaries between land and water bodies. Similar to the patterns observed in Fig. 8, the linear kernel-driven and DCF models produce some outliers characterized by pronounced overestimations, whereas the bilinear interpolation, RF-based kernel-driven, and RF-based DCF models exhibit reduced thermal contrast. PGDM yields better downscaling results that are more consistent with the ground truth, achieving an RMSE of 1.36 K and an SSIM of 0.90.

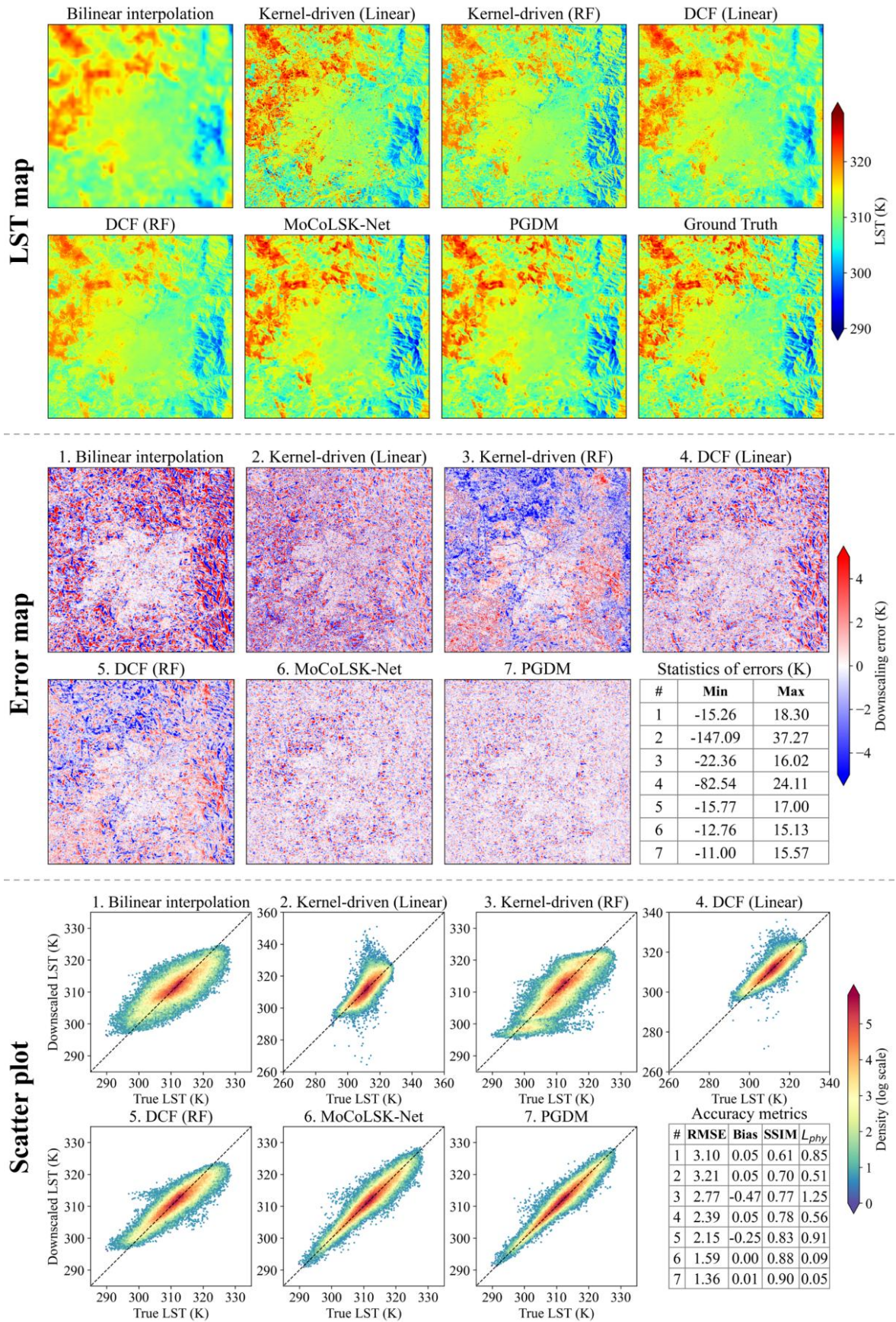

**Fig. 8.** Comparison of 10× downscaling results from all seven algorithms over Santiago, Chile in the Landsat_GLB dataset.

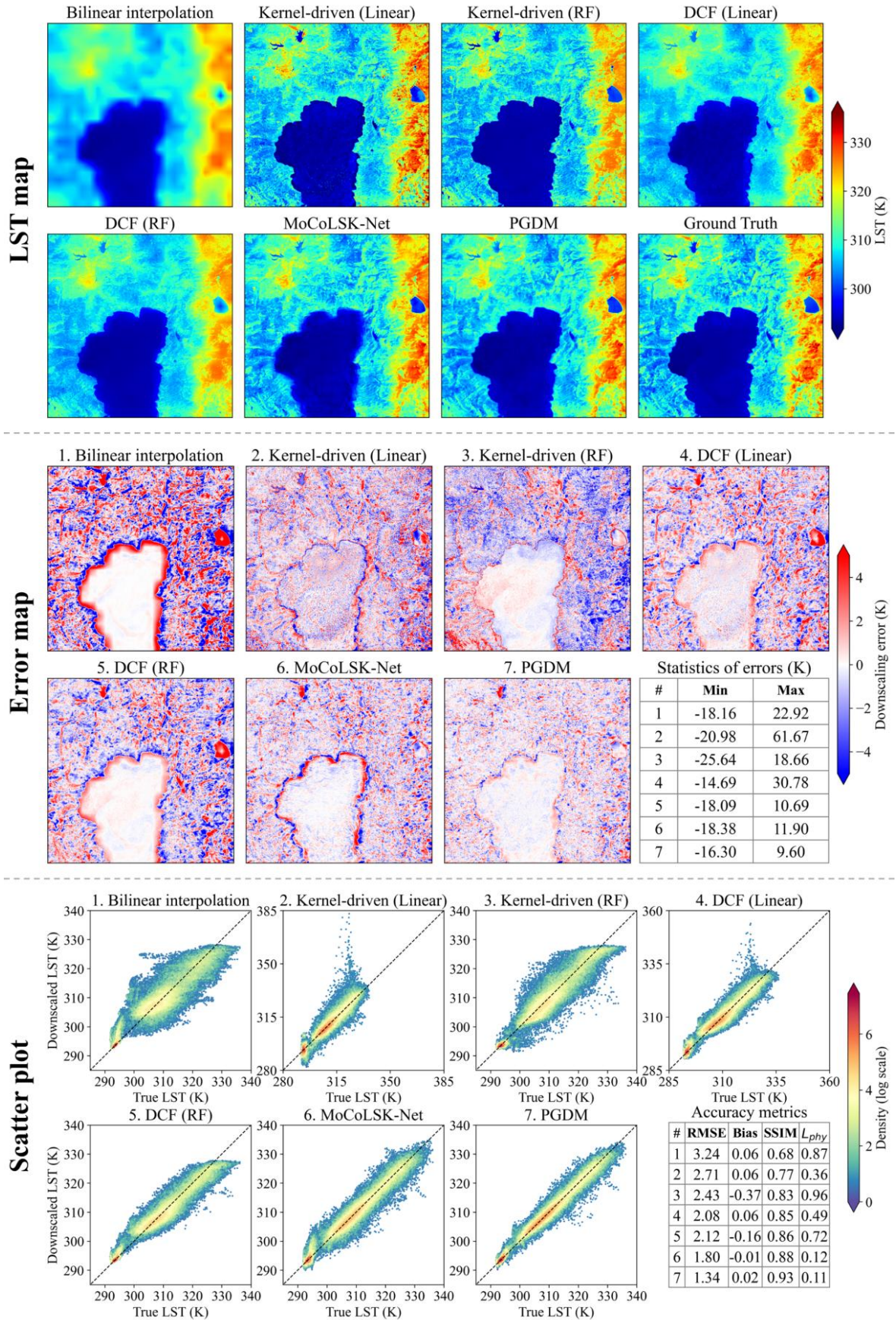

**Fig. 9.** Comparison of 20× downscaling results from all seven algorithms over Lake Tahoe, America in the Landsat_GLB dataset.

### 3.3 Performances on the ASTER_GLB dataset

The seven downscaling methods were finally applied on the ASTER_GLB dataset to test the stability of deep learning models to different sensors. Considering the differences in sensor characteristics, channel configurations, and LST retrieval algorithms, LST estimates from Landsat and ASTER are expected to exhibit systematic discrepancies. However, in this study, the deep learning models trained on the Landsat dataset are directly applied to downscale ASTER-simulated LST without correcting for these systematic errors, due to the limited number of synchronized observations available between these two sensors. Consequently, the performance of the deep learning models may deteriorate because of data drift. In contrast, kernel-driven methods, which establish the relationship between LR ASTER LST and kernels for each individual scene during downscaling, are less affected by systematic errors.

Table 7 summarized the performances of different downscaling methods on the ASTER_GLB dataset for both 10× and 20× downscaling. Under the unfavorable condition of missing systematic error correction, the deep learning models still achieve better downscaling results than the other methods. Specifically, the RMSE, SSIM, and $\mathcal{L}_{phy}$ of PGDM for 10× (20×) downscaling are 1.764 (1.930) K, 0.849 (0.833), and 0.042 (0.067) K, respectively, representing the best performance among all the seven methods. Similar to the results in Sections 3.1 and 3.2, the bilinear interpolation, kernel-driven, and DCF methods generally suffer from larger downscaling uncertainty and fail to satisfy the physical consistency during downscaling.

**Table 7.** Performances of different LST downscaling methods over the ASTER_GLB dataset.

| Method | For 10× / 20× downscaling | | | |
|---|---|---|---|---|
| | RMSE (K) | Bias (K) | SSIM | $\mathcal{L}_{phy}$ (K) |
| Bilinear | 2.378 / 2.780 | 0.029 / 0.039 | 0.748 / 0.728 | 0.622 / 0.627 |
| Kernel-driven (Linear) | 2.507 / 2.731 | 0.029 / 0.039 | 0.760 / 0.742 | 0.354 / 0.293 |
| Kernel-driven (RF) | 2.484 / 2.589 | -0.071 / -0.082 | 0.766 / 0.763 | 0.843 / 0.808 |
| DCF (Linear) | 2.079 / 2.275 | 0.029 / 0.039 | 0.801 / 0.790 | 0.387 / 0.378 |
| DCF (RF) | 2.112 / 2.330 | -0.028 / -0.021 | 0.802 / 0.790 | 0.631 / 0.600 |
| MoCoLSK-Net | 1.870 / 2.036 | 0.002 / 0.006 | 0.837 / 0.821 | 0.064 / 0.070 |
| PGDM (Proposed) | 1.764 / 1.930 | 0.012 / 0.021 | 0.849 / 0.833 | 0.042 / 0.067 |

Fig. 10 compares the 20× downscaling results from all methods over Gaigoab, Namibia. Regarding the downscaled LST maps, only PGDM effectively reconstructs the low-temperature details associated with complex terrains. As shown in the error maps, PGDM exhibits generally lower error magnitudes, characterized by lighter color regions across the study area. In contrast, the other downscaling methods demonstrate larger and more spatially heterogeneous errors. In the scatter plot, the downscaled LSTs from PGDM align more closely with the 1:1 line, with the lowest RMSE of 1.41 K and highest SSIM of 0.88. The results from RF-based kernel-driven and DCF methods as well as bilinear interpolation suffer from reduced thermal contrast, while the linear kernel-driven and DCF methods again produce some outliers and noise.

Fig. 11 further presents the evaluation results of 10× downscaling over Yueyang, China. In this case, the two deep learning models demonstrate comparably stable performance, achieving RMSE values below 1.3 K and SSIM around 0.89. In comparison, the error maps of the other five methods appear more spatially heterogeneous, particularly showing apparent overestimation in river areas. The patterns observed in the scatter plot are consistent with those shown in Figs. 8–10 and are not reiterated here.

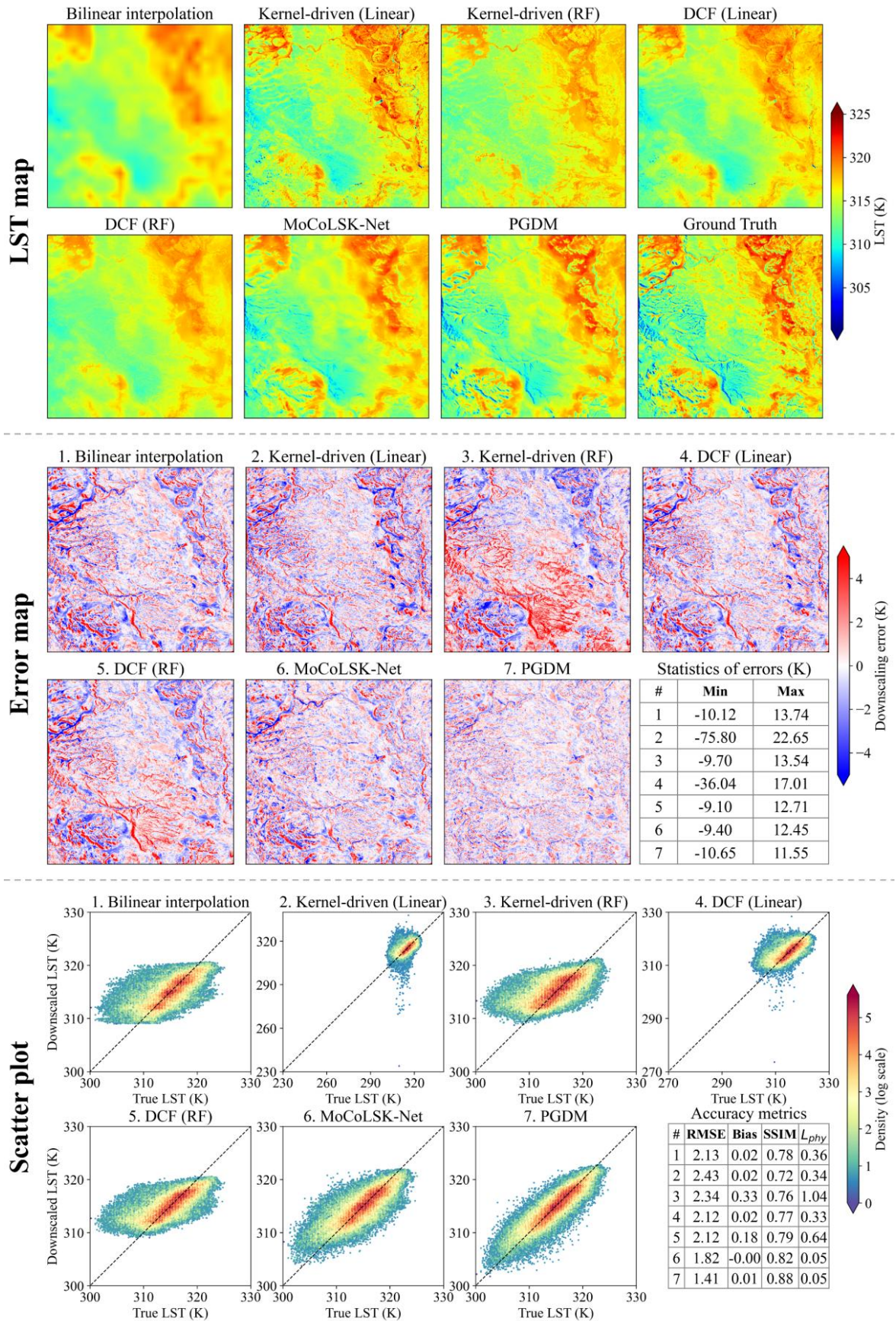

**Fig. 10.** Comparison of 20× downscaling results from all seven algorithms over Gaigoab, Namibia in the ASTER_GLB dataset.

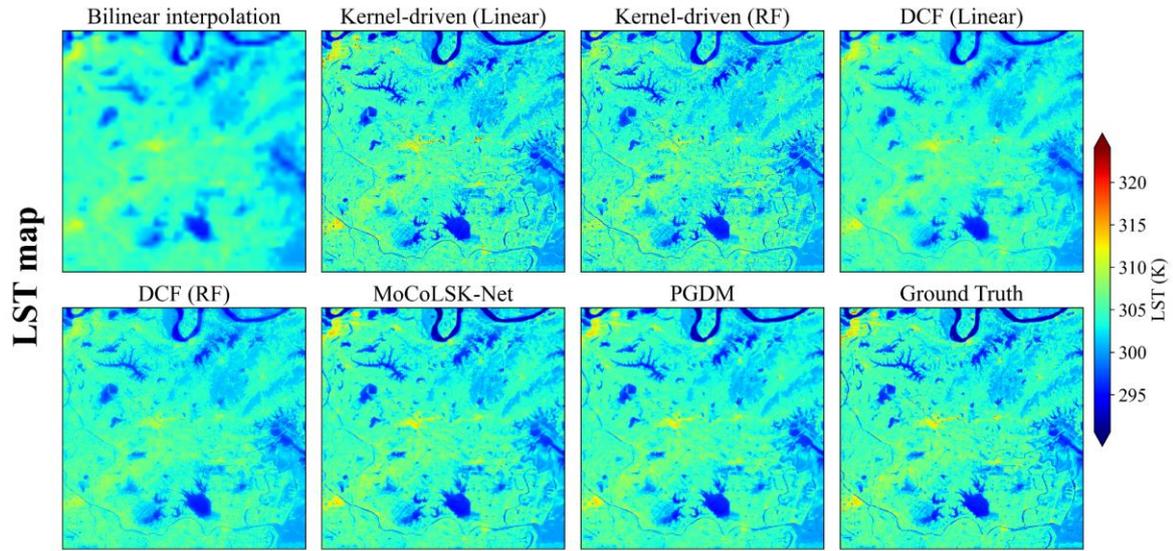
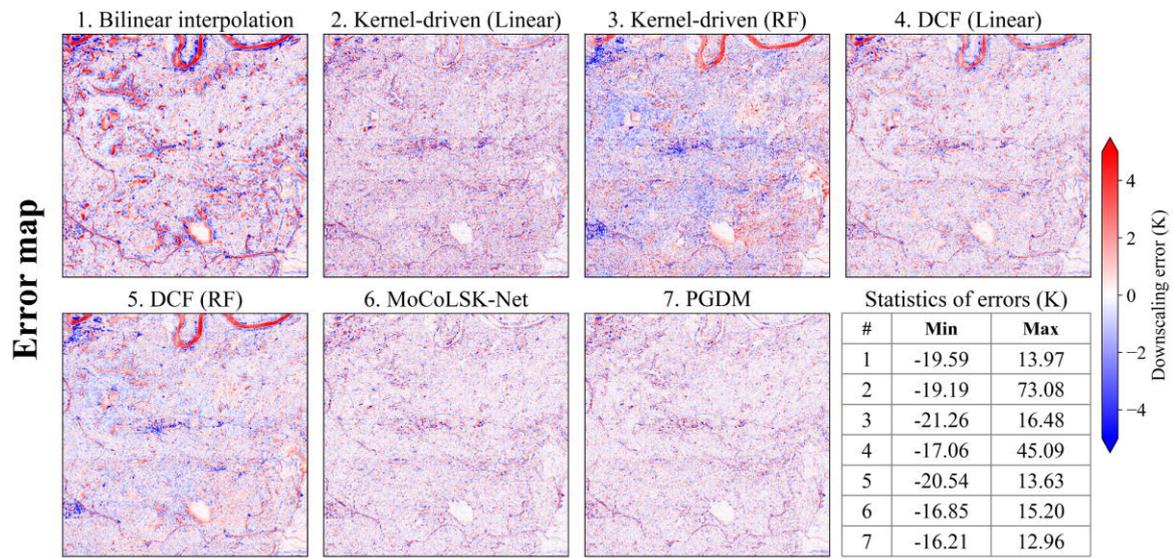
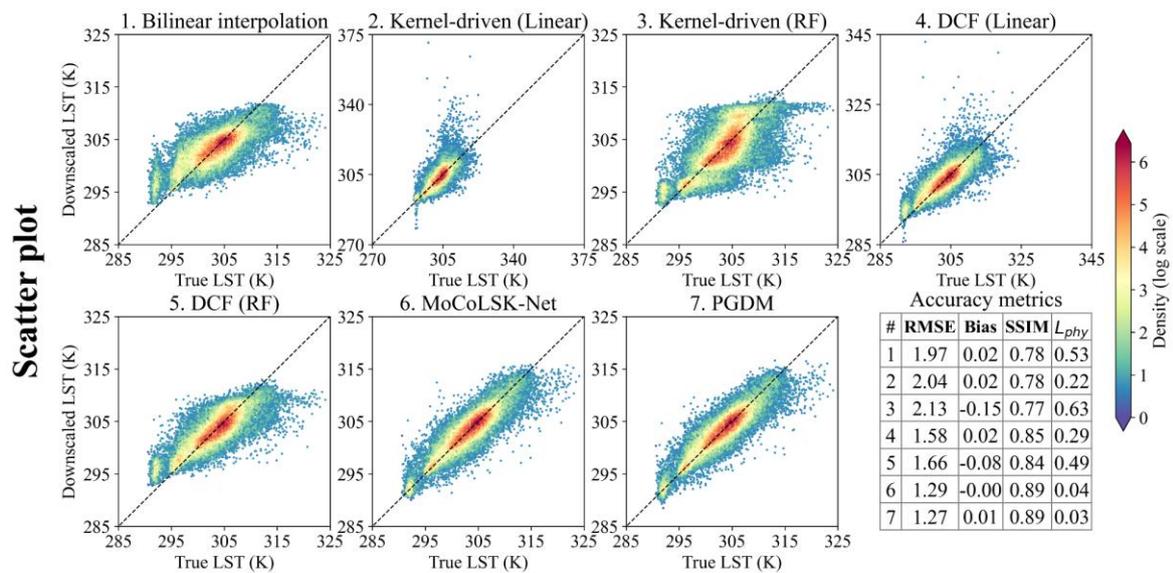

**Fig. 11.** Comparison of 10× downscaling results from all seven algorithms over Yueyang, China in the ASTER_GLB dataset.

### 3.4 Scene-level uncertainty indicator

In contrast to conventional downscaling methods that establish a deterministic mapping between $T_s^{HR}$ and $T_s^{LR}$, PGDM is essentially a generative model, which can produce diverse downscaling results driven by stochastic noise during the denoising process. This stochastic nature allows for ensembling within a single diffusion model: by initializing different random seeds to govern noise generation, multiple realizations can be generated and subsequently aggregated to provide useful information.

In contrast to Sections 3.1–3.3, where PGDM was executed only once for simplicity, this section performs five independent runs of PGDM on the Landsat_CN20 test set, with the mean and STD values calculated for each pixel. Only 10× downscaling was analyzed for simplicity. The mean values represent the ensembled downscaling results, leading to a slight reduction in downscaling RMSE from 0.610 K to 0.585 K compared with the single-run result in Section 3.1. Although the accuracy improvement is limited, the primary significance of the ensemble lies in the inherent uncertainty quantified by the STD values.

To mitigate the localized and irregular noise in data, the pixel-level diffusion STD and absolute error were averaged for each scene. As shown in Fig. 12(a), the scene-level diffusion STD exhibits a strong positive correlation with the corresponding mean absolute error. The Pearson correlation coefficient (r) reaches 0.971, and the majority of data points are tightly clustered around the linear regression line. It should be noted that the diffusion STD values are relatively small, as the solution space of PGDM is tightly constrained by the prior geophysical guidance. In comparison, the relationship between scene-level downscaling error and the STD of LST distribution in Fig. 12(b) appears more dispersed. This is reasonable, as the STD of LST distribution serves only as an external measure of task complexity and does not exhibit a definitive correlation with the actual downscaling error. In contrast, the STD derived from PGDM's diverse outputs reflects the inherent uncertainty associated with the model's decision-making process and therefore shows a stronger relationship with the final downscaling error. Consequently, the mean diffusion STD can be used to calibrate scene-level downscaling error in practice, which represents an advantage of PGDM over the other downscaling methods.

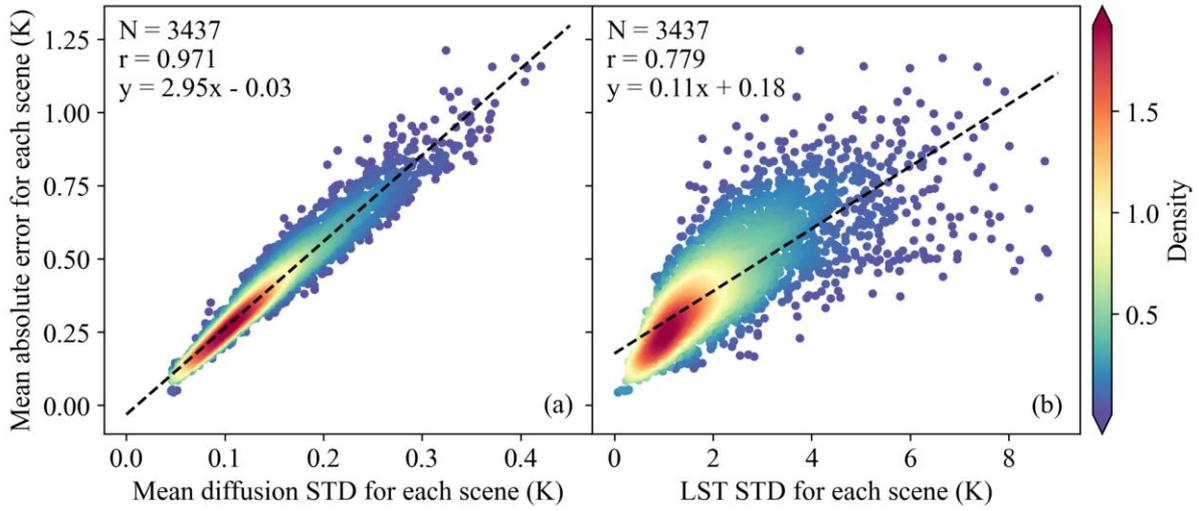

**Fig. 12.** Scene-level relationship between mean absolute error and (a) the STD of LST, and (b) the mean diffusion STD in the Landsat_CN20 test set for 10× downscaling.

## 4. Discussion

### 4.1 Upscaling methods of LST

For the "upscaling-downscaling" evaluation strategy, two common approaches are used for LST upscaling: the simple averaging method and the energy conservation method without considering $\varepsilon_{bb}$. However, a comprehensive analysis of upscaling errors under varying conditions is still lacking, as previous studies have typically been limited to small datasets (Hu et al., 2015; Liu et al., 2006). By taking Eq. (5) with the inclusion of $\varepsilon_{bb}$ as the reference truth, errors of these two upscaling methods were assessed by a comprehensive simulation analysis. Compared to the reference truth, the first method neglects both emissivity and exponent, while the second method ignores emissivity alone. For simplicity, this study assumes a flat surface without terrain effects, a common assumption adopted in most downscaling studies.

HR LST samples following the Gaussian distribution with varying mean and STD values were generated to simulate the thermal distribution within a single LR pixel. The mean values range from 200 K to 340 K with a step of 10 K, while STD range from 2 K to 10 K with a step of 2 K. For $\varepsilon_{bb}$, two scenarios were simulated. The first one assumed that all pixels shared the same $\varepsilon_{bb}$ value, which represented an ideal homogeneous scenario. The second one followed a uniform distribution between 0.8 and 1.0, simulating an extremely heterogeneous situation. Finally, a total of 150 distinct cases were simulated. For each case, errors were computed 10 times using different random seeds, and the final results were obtained by averaging these runs. The number of HR pixels was set to 400 here, while we found the analyzing results were robust to pixel numbers.

Fig. 13 provides the assessment results for two upscaling methods. Only 75 cases under the second scenario of $\varepsilon_{bb}$ are presented, as the discrepancies between these two $\varepsilon_{bb}$ scenarios are very minimal. In other words, the influence of $\varepsilon_{bb}$ in upscaling LST is generally negligible, resulting in the stable and promising performance of the second upscaling method. In contrast, further neglecting the exponent leads to more obvious accuracy loss in the first simple averaging method, particularly under spatially heterogeneous LST conditions. Therefore, the energy conservation method is preferable for simulating LR LSTs.

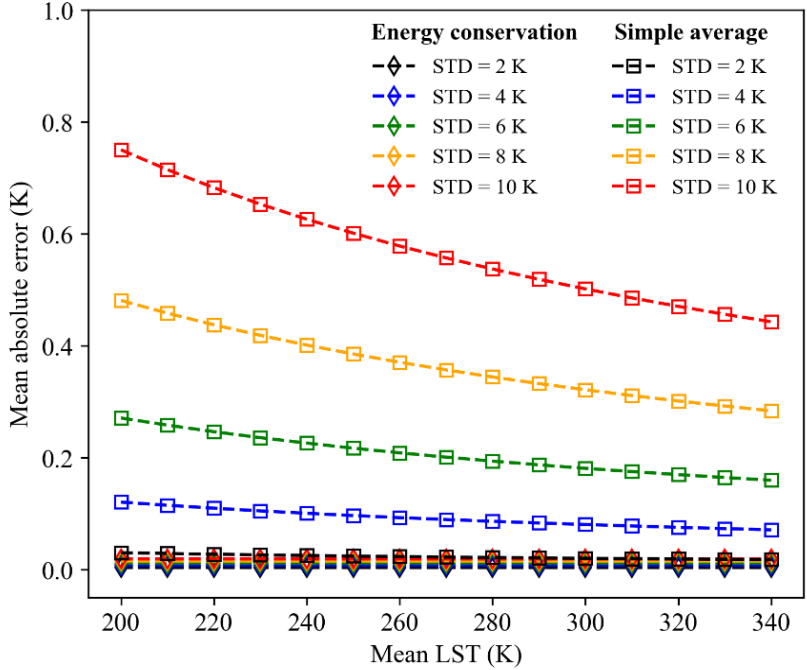

**Fig. 13.** Accuracy loss of the simple averaging and energy conservation methods without emissivity for LST upscaling under different simulated cases. The energy conservation method considering emissivity is regarded as the reference for calculating the mean absolute error.

### 4.2 Ablation experiments

The denoising neural network in PGDM typically employs well-established architecture (e.g., encoder-decoder) and modules (e.g., ResBlocks and non-local networks) that are widely accepted in the research community. We intentionally avoided incorporating overly complex architectures or modules to gain minor accuracy improvements, as LST downscaling is not simply a task of chasing benchmark performance. Accordingly, the ablation experiments here primarily examine the effects of $C_{base}$ on the final downscaling results produced by PGDM.

Table 8 compares the influence of $C_{base}$ on the downscaling accuracy and computational complexity of PGDM, in which the Landsat_CN20 test set for 10× downscaling is used for assessment. Increasing the number of channels leads to a decrease in RMSE, but at the cost of higher computational complexity. Furthermore, the marginal improvement in accuracy from increasing $C_{base}$ becomes progressively smaller. To achieve a trade-off between performance and computational efficiency, $C_{base}$ was finally set to 32 in this study.

**Table 8.** RMSE and computational complexity for PGDM with different $C_{base}$ values on the Landsat_CN20 test set for 10× downscaling.

| $C_{base}$ | RMSE (K) | Parameters | FLOPs |
|---|---|---|---|
| 8 | 0.771 | 892.54K | 3.70G |
| 16 | 0.658 | 3.18M | 14.58G |
| 32 | 0.610 | 12.17M | 57.86G |
| 64 | 0.574 | 47.72M | 230.55G |

### 4.3 Limitations and further improvements

In this study, we adopted the widely used "upscaling-downscaling" strategy to evaluate the performance of downscaling methods. The T-based validation was not introduced, as the uncertainties from the downscaling algorithms themselves, LR LST observations, in-situ measurements, and other factors could hinder an independent and reliable assessment of model performance (Dong et al., 2023; Xia et al., 2019a). We plan to integrate the proposed PGDM with our previous clear-sky and all-sky LST retrieval algorithms (Cheng et al., 2025; Zhang et al., 2025a; Zhang et al., 2025b; Zhang et al., 2024) to form a complete framework for LST retrieval and post-processing, which is expected to produce all-sky LST information with both high spatial and temporal resolutions. Subsequently, the T-based validation will be employed to assess the absolute accuracy of the overall framework.

Several representative downscaling methods were reproduced for comparison over a large spatial extent, including interpolation-based methods, kernel-driven models, hybrid methods combining kernel-driven and interpolation, and deep learning models. The other downscaling methods (e.g., weight-function fusion, thermal mixing, and physical SEB) were not included due to the difficulty of obtaining the required input parameters or the high computational complexity. In addition, both the weight-function fusion and thermal mixing methods appear more empirical, which cannot fully satisfy the paradigm derived from the SEB-based geophysical reasoning, reducing their comparability with the other downscaling methods.

PGDM was trained on the Landsat_GLB dataset and directly applied to the ASTER_GLB dataset without correcting for systematic errors, owing to the limited synchronized observations between Landsat and ASTER. Although potentially constrained by the "data shift" problem, PGDM still yielded better downscaling results compared to the other methods. The ECOsystem Spaceborne Thermal Radiometer Experiment on Space Station (ECOSTRESS) may bring synchronized HR observations with Landsat due to its irregular orbit (Fisher et al., 2020; Zhang et al., 2025c), while it has been reported to suffer from striping and other issues (Kustura et al., 2025). Therefore, the ECOSTRESS data was not used in this study. In the future, when applying the PGDM model to real LST products with higher temporal resolutions (such as geostationary satellites), the correction of systematics errors may be easier.

Finally, the benefits of downscaled LST with high spatiotemporal resolutions also merit further investigation in agriculture and ecology, such as field-scale drought monitoring (Jia et al., 2025) and for accurately capturing the diurnal dynamics of ecosystem processes (Xiao et al., 2021).

## 5. Conclusion

Despite remarkable advancements have been achieved in LST downscaling, several common challenges remain, including the lack of comprehensive benchmark datasets, the difficulty of efficient application at large scales, and the inability of models to self-assess the uncertainty inherent in the downscaling process. To address these issues, this study focused on the following three aspects.

First, three comprehensive datasets were constructed, including the Landsat_CN20 dataset containing 22,909 image patches across mainland China, and the Landsat_GLB and ASTER_GLB datasets encompassing 40 heterogeneous regions worldwide. These datasets provide a comprehensive benchmark for LST downscaling model training and evaluation, thereby facilitating future studies in this field.

Next, guided by the SEB-based geophysical reasoning, we proposed PGDM for LST downscaling. Conditioned on LR LST and HR auxiliary geophysical parameters, PGDM samples from the complex conditional posterior distribution associated with LST downscaling to generate HR LST predictions. Comprehensive evaluations demonstrated that PGDM delivered high-quality downscaling results with both computational efficiency and robustness. PGDM was also adaptive to different combinations of HR auxiliary parameters, making it more flexible.

Finally, the inherent stochasticity in the denoising process of PGDM enables the model ensembling, and the derived scene-level mean diffusion STD was found to exhibit a strong positive linear relationship with the scene-level mean absolute downscaling error. This capability distinguishes PGDM from the other deterministic downscaling approaches, which allows PGDM to self-assess its downscaling results and thereby provide valuable guidance for downstream applications.

Given its advantages, PGDM has the potential to serve as a reliable solution for LST downscaling. In the future study, we intend to integrate PGDM with our existing clear-sky and all-sky LST algorithms, establishing a unified framework capable of producing seamless LST estimates with both high spatial and temporal resolutions.

## Acknowledgements

This work was supported in part by the China Scholarship Council (CSC), in part by the National Natural Science Foundation of China under Grant 42230109, in part by the Yunling Scholar Project of the "Xingdian Talent Support Program" of Yunnan Province under Grant 41961053, in part by the Yunnan International Joint Laboratory for Integrated Sky-Ground Intelligent Monitoring of Mountain Hazards under Grant 202403AP140002, in part by the Platform Construction Project of High-Level Talent in the Kunming University of Science and Technology (KUST) under Grant 141120210012, and in part by the Luxembourg Fonds National de la Recherche (FNR) CORE programme (ENCORE, C23/SC/18171206). The entire Landsat 8 L2C2T1 images for constructing the Landsat_CN20 dataset were downloaded from EarthExplorer (https://earthexplorer.usgs.gov/). The AST_08 dataset was acquired through Earthdata Search (https://search.earthdata.nasa.gov/). Other datasets including the regional Landsat 8 L2C2T1 images in the Landsat_GLB dataset, HLS, Copernicus GLO-30 DEM and CGLS-LC, were preprocessed and obtained from the Google Earth Engine (https://earthengine.google.com/).

# Supplementary materials for "PGDM: Physically guided diffusion model for land surface temperature downscaling"

Huanyu Zhang[a,c,d], Bo-Hui Tang[b,e,f,a,*], Tian Hu[d,**], Yun Jiang[a], Zhao-Liang Li[g,a]

[a] State Key Laboratory of Resources and Environmental Information System, Institute of Geographic Sciences and Natural Resources Research, Chinese Academy of Sciences, Beijing 100101, China

[b] Faculty of Land Resource Engineering, Kunming University of Science and Technology, Kunming 650093, China

[c] College of Resources and Environment, University of Chinese Academy of Sciences, Beijing 100049, China

[d] Luxembourg Institute of Science and Technology, Belvaux 4362, Luxembourg

[e] Yunnan Key Laboratory of Quantitative Remote Sensing, Kunming 650093, China

[f] Yunnan International Joint Laboratory for Integrated Sky-Ground Intelligent Monitoring of Mountain Hazards, Kunming 650093, China

[g] State Key Laboratory of Efficient Utilization of Arid and Semi-arid Arable Land in Northern China, Institute of Agricultural Resources and Regional Planning, Chinese Academy of Agricultural Sciences, Beijing 100081, China

* Corresponding author at: Faculty of Land Resource Engineering, Kunming University of Science and Technology, Kunming, 650093, China.
** Corresponding author at: Luxembourg Institute of Science and Technology, Belvaux 4362, Luxembourg.
*E-mail addresses:* tangbh@kust.edu.cn (B.-H. Tang), tian.hu@list.lu (T. Hu)

**This document includes:**
Supporting Text S1
Supporting Figure S1
Supporting Tables S1 and S2

**Text S1.** Detailed methods for compiling the three datasets

For compiling the Landsat_CN20 dataset, all Landsat 8 L2C2T1 images covering mainland China in 2020 with cloud coverage below 5% and water coverage below 40% were downloaded at first, which provided $T_s^{HR}$ and $\rho^{HR}$. For the convenience of model training, each image was cropped into non-overlapping patches of 160×160 pixels, and only those entirely composed of clear-sky and valid pixels were retained. To ensure a relatively balanced spatial and temporal distribution of the dataset, up to 2,000 patches were sampled per month using a weighted random sampling algorithm (Efraimidis and Spirakis, 2006), with weights assigned inversely to the patch numbers in each image. For the selected patches, $LULC^{HR}$ and $H^{HR}$ were downloaded from the Copernicus and CGLS-LC100 in 2019 and GLO-30 DEM datasets, respectively, and $NDXI^{HR}$ was calculated from $\rho^{HR}$. Based on the above methods, the Landsat_CN20 dataset with 22,909 image patches was finally compiled. Fig. S1 provides an example in the Landsat_CN20 dataset.

The Landsat_GLB dataset was also compiled from the Landsat 8 L2C2T1, Copernicus GLO-30 DEM, and CGLS-LC100 data. For each selected study region, all corresponding Landsat images between 2014 and 2024 were filtered, and the scene with the highest ratio of valid pixels (typically equal to or very close to 100%) was selected. Missing values, when present, were filled using the "focalMean" function in the Google Earth Engine (GEE). Table S1 provides specific information about the evaluation regions in the Landsat_GLB dataset, including their geolocations, land cover types, and corresponding Landsat images.

The ASTER_GLB dataset encompassed 20 additional heterogeneous regions worldwide, with one completely clear-sky ASTER LST image selected for each region. Given that surface reflectance generally changes slowly over time, a 30-day temporal window was constructed centered on the acquisition date of each ASTER LST image. If a completely clear HLS image was available within this window, the one with the shortest temporal difference was selected; otherwise, a monthly median composite was generated on GEE. The basic information of these evaluation regions in the ASTER_GLB dataset is summarized in Table S2.

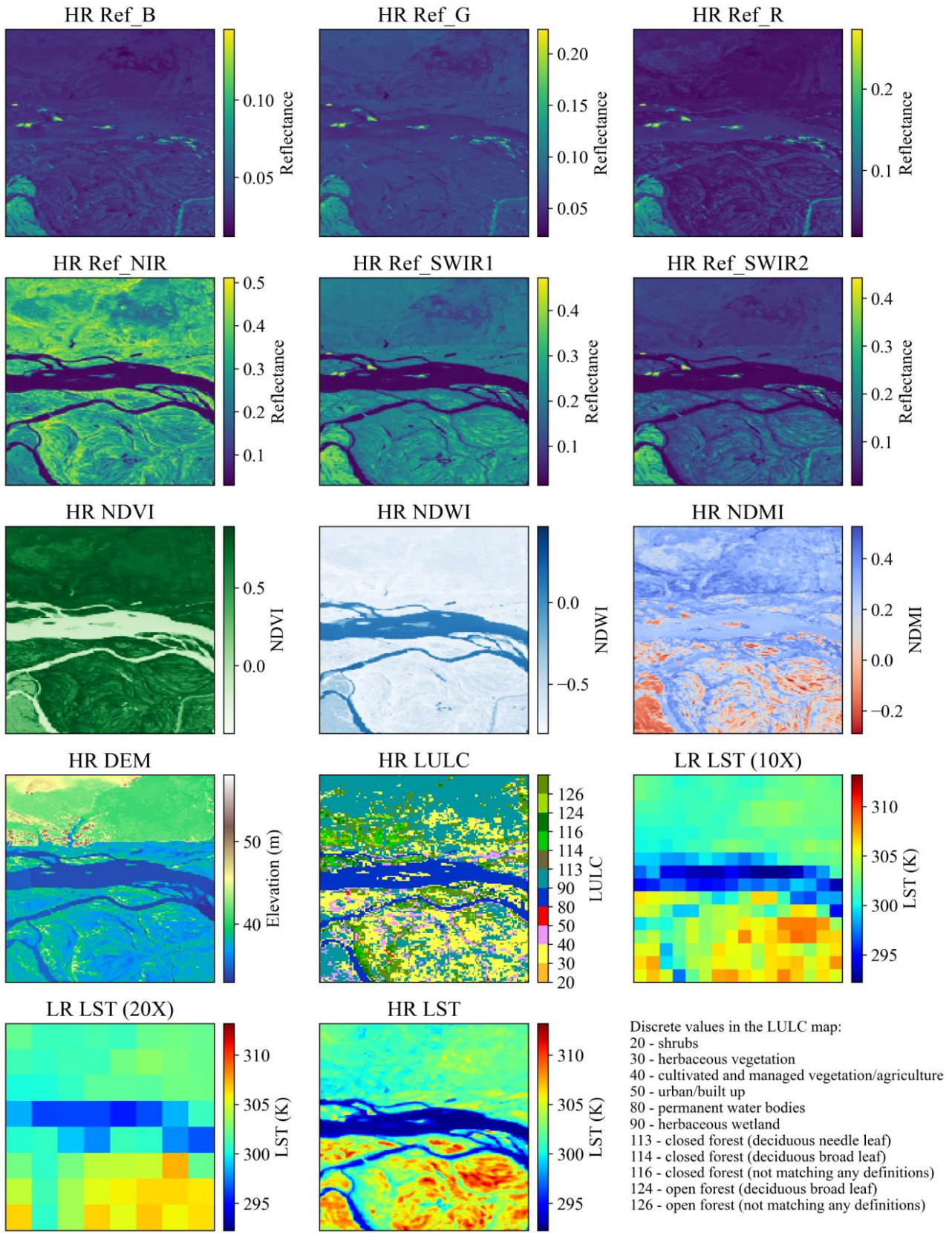

**Fig. S1.** An example data pair from the Landsat_CN20 dataset.

**Table S1.** Global test regions outside China in the Landsat_GLB dataset. The frequency distribution of land cover types for each area was derived from the MCD12Q1 dataset, with the modal value over 2014–2024 used as the representative class for each pixel. The Landsat data are named following the "LC08_PathRow_yyyymmdd" format.

| Area | Central Lat, Lon | Land cover type | Landsat data |
| --- | --- | --- | --- |
| Paris, France | 48.85, 2.35 | Urban (71.2%) | LC08_199026_20200807 |
| Alps, Switzerland | 46.1, 9.0 | Deciduous broadleaf forests (40.4%), Woody savannas (23.5%) | LC08_194028_20220709 |
| Roma, Italy | 42.043, 12.543 | Savannas (30.9%) Cropland (29.7%), Urban (18.6%) | LC08_191031_20150717 |
| Zabuze, Poland | 52.3, 23.0 | Cropland (47.1%), Mixed forests (18.9%) | LC08_186024_20210119 |
| Typa, Russia | 64.3, 100.2 | Savannas (97.7%) | LC08_141015_20160518 |
| Lake Tahoe, America | 39.2, -120.0 | Grassland (24.4%), Savannas (22.7%), Water bodies (20.9%) | LC08_043033_20220723 |
| Phoenix, America | 33.45, -112.1 | Urban (63.6%), Open shrublands (21.0%) | LC08_037037_20140520 |
| Louisiana, America | 32.2, -91.5 | Cropland (52.2%), Deciduous broadleaf forests (13.8%) | LC08_023038_20160507 |
| Alberta, Canada | 57.4, -111.0 | Woody savannas (61.2%), Evergreen needleleaf forests (26.4%) | LC08_042020_20150526 |
| Santiago, Chile | -33.45, -70.65 | Urban (32.1%), Savannas (30.5%), Grassland (30.0%) | LC08_233083_20230219 |
| Maranhao, Brazil | -4.3, -43.88 | Woody savannas (65.4%), Evergreen broadleaf forests (18.9%) | LC08_220063_20190909 |
| Mount Kenya, Kenya | 0.4, 37.3 | Grassland (96.7%) | LC08_168060_20140203 |
| Qaryat, Libya | 30.5, 13.5 | Barren (99.8%) | LC08_188039_20140215 |
| Zongo, DR Congo | 4.35, 18.6 | Savannas (55.9%), Evergreen broadleaf forests (27.1%) | LC08_181057_20141129 |
| Karasu, Kazakhstan | 48.85, 54.7 | Grassland (96.5%) | LC08_165026_20140724 |
| Tamil Nadu, India | 11.35, 78.5 | Cropland (61.0%), Savannas (15.0%) | LC08_143052_20210207 |
| Jalod Block, India | 24.5, 75.35 | Cropland (46.1%), Grassland (42.4%) | LC08_147043_20140131 |
| Bandung, Indonesia | -7.0, 107.5 | Natural vegetation mosaics (25.2%), Evergreen broadleaf forests (22.6%), Woody savannas (20.7%) | LC08_122065_20190911 |
| Lake Mackay, Australia | -22.3, 128.7 | Barren (49.7%), Open shrublands (49.5%) | LC08_106075_20140201 |
| Warialda, Australia | -29.35, 150.35 | Grassland (69.6%) | LC08_091080_20140207 |

**Table S2.** Global test regions in the ASTER_GLB dataset. The frequency distribution of land cover types for each area was derived from the MCD12Q1 dataset, with the modal value over 2014–2024 used as the representative class for each pixel. The ASTER data are named following the "AST_08_004mmddyyyyHHMMSS" format.

| Area | Central Lat, Lon | Land cover type | ASTER data |
| --- | --- | --- | --- |
| Beijing, China | 39.781, 116.416 | Urban (70.3%), Cropland (27.6%) | AST_08_00410052020031711 |
| Yueyang, China | 29.517, 112.615 | Cropland (49.1%), Savannas (19.8%) | AST_08_00404292017031407 |
| Harbin, China | 45.901, 128.597 | Cropland (59.7%), Deciduous broadleaf forests (24.8%) | AST_08_00409252019022557 |
| Yunnan, China | 39.256, 76.053 | Savannas (37.0%), Woody savannas (20.4%), Urban (15.5%) | AST_08_00411252020035204 |
| Kashgar, China | 25.202, 102.673 | Cropland (60.5%), Grassland (25.2%) | AST_08_00407262021055452 |
| Madrid, Spain | 40.306, -3.91 | Grassland (42.8%), Cropland (25.5%), Urban (16.8%) | AST_08_00407032020111907 |
| Mühltroff, Germany | 50.559, 11.947 | Cropland (34.7%), Savannas (18.7%), Grassland (17.4%) | AST_08_00406022020102041 |
| Belarus | 54.252, 30.34 | Cropland (73.2%) | AST_08_00407262015091258 |
| Ryazan Oblast, Russia | 54.944, 42.424 | Mixed forests (50.4%), Woody savannas (23.9%) | AST_08_00408222017083451 |
| Cherkasy, Ukraine | 49.089, 32.196 | Cropland (67.2%) | AST_08_00410012018085547 |
| Kamloops, Canada | 50.462, -120.264 | Woody savannas (46.9%), Grassland (31.5%), Evergreen needleleaf forests (16.1%) | AST_08_00407252018191345 |
| Columbia, America | 34.112, -81.019 | Woody savannas (51.7%), Deciduous broadleaf forests (18.4%), Urban (15.3%) | AST_08_00405162014161808 |
| Missouri River, America | 44.847, -100.444 | Grassland (47.2%), Cropland (37.8%) | AST_08_00407132017174123 |
| Mexico City, Mexico | 19.26, -99.055 | Urban (36.9%), Cropland (20.1%), Grassland (17.2%) | AST_08_00402092020171739 |
| Carnamah, Australia | -29.753, 116.009 | Cropland (65.7%), Grassland (24.5%) | AST_08_00411132018022258 |
| Santa Fe, Argentina | -28.769, -60.869 | Grassland (79.7%), Savannas (19.5%) | AST_08_00411262021140718 |
| Amazonas, Venezuela | 4.967, -66.149 | Evergreen broadleaf forests (89.4%) | AST_08_00403042020145333 |
| Gaigoab, Namibia | -27.222, 17.125 | Open Shrublands (57.8%), Barren (37.9%) | AST_08_00410062019090309 |
| Cairo, Egypt | 29.793, 31.092 | Barren (67.2%), Urban (22.4%) | AST_08_00404182015084751 |
| Jofeyr, Iran | 31.119, 48.093 | Barren (64.7%), Grassland (23.9%) | AST_08_00410092016073323 |